\documentclass[]{spie}  

\usepackage{graphicx}
\usepackage{amsmath}
\usepackage{subfig}
\usepackage{multirow}
\usepackage{epstopdf}
\usepackage{booktabs}
\usepackage{float}
\usepackage{url}

\graphicspath{{assets/}}

\title{Prospects for exoplanet imaging in multi-star systems with starshades}

\author{Dan Sirbu\supit{1}, Ruslan Belikov\supit{1}, Eduardo Bendek\supit{1}, Elias Holte\supit{1,2}, A J Eldorado Riggs\supit{3}, Stuart Shaklan\supit{3}
\skiplinehalf
\supit{1}
NASA Ames Research Center, Moffett Field, Mountain View, CA
\skiplinehalf
\supit{2}
Minnesota State University Moorhead, Moorhead, CA
\skiplinehalf
\supit{3}
Jet Propulsion Laboratory, California Institute of Technology, Pasadena, CA}

\begin{document}
 
  \maketitle 

\begin{abstract}

We explore the capabilities of a starshade mission to directly image multi-star systems. In addition to the diffracted and scattered light for the on-axis star, a multi-star system features additional starlight leakage from the off-axis star that must also be controlled. A basic option is for additional starshades to block the off-axis stars. An interesting option takes the form of hybrid operation of a starshade in conjunction with an internal starlight suppression. Two hybrid scenarios are considered. One such scenario includes the coronagraph instrument blocking the on-axis star, with the starshade blocking off-axis starlight. Another scenario uses the wavefront control system in the coronagraph instrument and using a recent Super-Nyquist Wavefront Control (SNWC) technique can remove the off-axis star’s leakage to enable a region of high-contrast around the on-axis star blocked by the starshade. We present simulation results relevant for the WFIRST telescope.

\end{abstract}

\keywords{High-Contrast Imaging, Binary Stars, External Occulters, Starshades, Wavefront Control, Super-Nyquist Wavefront Control, Coronagraphs}

\section{Introduction} 

The main technical challenge associated with direct imaging of exoplanets is the control of diffracted and scattered light from the host star so that a dim planetary companion can be imaged at small angular separations and at high contrast ratios\cite{DesMarais02}. Upcoming space missions including Wide Field Infrared Infrared Survey Telescope  (WFIRST) \cite{Macintosh17}, Habitable Exoplanet Imaging Mission (HabEx) \cite{Menesson16, Stahl17}, and Large UV/Optical/Infrared Surveyor (LUVOIR) \cite{Bolcar17} are currently exploring the usage of both internal coronagraphs and starshades to enable direct imaging of exoplanets. Multi-star systems pose additional challenges for both types of starlight suppression systems due to the combined off-axis star's diffraction and aberration leakage into the region of interest. As a result, many multi-star systems are considered unsuitable for direct imaging and not included in direct imaging survey target lists, even though the majority of Sun-like (FGK) stars belong to multistar systems. Alpha Centauri, the nearest star system to the Sun, is an example of a star system excluded from high-contrast imaging surveys since it is a binary system.

A \emph{starshade}, also called an external occulter, is a specially-shaped spacecraft flying in formation with a space telescope. The occulter is placed along the line-of-sight of the telescope and the (on-axis) target star. The starshade blocks most of the stellar flux before reaching the telescope pupil allowing for the high-contrast imaging of exoplanets. The edge of the starshade  has to be optimally designed to minimize diffraction and create a shadow with deep suppression over a region that will cover the telescope's diameter such that it attains the necessary $10^{-10}$ (or better) contrast for imaging and characterizing an Earth-like planet. 

A \emph{coronagraph} consists of a set of internal optics that apply phase and amplitude changes at the image and pupil planes of the telescope to enable high-contrast imaging. Whereas coronagraph designs are theoretically capable of achieving contrast ratios necessary for imaging Earth-like planets, in practice wavefront errors introduced by surface aberrations of the optics degrade performance to orders of magnitude above
the coronagraph's design by leakage of the stellar point spread function (PSF). To recover a region of high-contrast (called the \emph{dark hole}) among a limited-contrast speckle-field created by the presence of optical aberrations, techniques from adaptive optics have been used in conjunction with deformable mirrors (DM) \cite{Malbet95}. Thus a coronagraph (at least the ones currently studied for upcoming space missions) will always include a wavefront control (WFC) loop using DMs to enable the creation of high-contrast dark holes.

Hybrid high-contrast imaging systems that use both starshades and coronagraphs simultaneously have been studied in the past with the goal of relaxing tolerances through their joint operation for the starlight suppression of single-star systems. It was found that amplitude variations in the shadow caused by the starshade upstream of the telescope pupil actually resulted in tighter coronagraph alignment requirements \cite{Cady08}; however, sequential operation of starshades and coronagraphs shows improved mission yields \cite{Stark16}. For example, the current HabEx mission architecture baselines both a starshade and coronagraph. A starshade rendezvous is currently being studied as a follow-on to the WFIRST mission which would already feature a coronagraph and integral field spectrograph (IFS) \cite{Mandell17}.

This study outlines the ways in which starshades can be utilized to enable high-contrast imaging of multi-star systems. Imaging multi-star systems introduces the additional challenge of suppressing the off-axis starlight from the additional stellar companion(s); the basic challenge and scientific utility of imaging multi-star systems is reviewed in Section \ref{sect:multiStarImaging}. Off-axis starlight leakage can be suppressed using either an off-axis starshade blocker or via wavefront control system that can remove off-axis speckles beyond the nominal control limit using Super-Nyquist Wavefront Control (SNWC) \cite{Thomas15} as discussed in Section \ref{sect:leakage}).

Thus, two starshades could enable high-contrast observation with one starshade operating conventionally on-axis and a second operating off-axis. A potentially more interesting option would be to use the starshade and coronagraph systems already baselined simultaneously to enable high-contrast imaging of multi-star systems. This could be achieved by either using the coronagraph and wavefront control system on-axis and blocking the off-axis star with the starshade. Alternatively, the on-axis star could be blocked by the starshade and the off-axis star's leakage could be removed using the coronagraph instrument's wavefront control system using SNWC. These scenarios are simulated for WFIRST for observations of the Alpha Centauri system in Section \ref{sect:wfirst}.

\section{Multi-Star Imaging} \label{sect:multiStar} \label{sect:multiStarImaging}

Direct imaging theoretically allows a complete census and follow-on characterization of a large number of nearby exoplanets. However, approximately 50\% of FGK stars within 20 pc and beyond are located in multi-star systems which present an additional imaging challenge due to the off-axis companion. Of particular interest to WFIRST science, out of the nearest 20 FGK stars, 13 are located within multi-star systems. Additionally, within 10 pc there are 70 known FGK stars out of which 42 stars are located in known (dynamical) multi-star systems. At further distances, binary companions pairs may not be as well characterized -- however, within 20 pc there are 510 FGK stars out of which 260 stars have known Washington Double Star (WDS) catalog entries \cite{Mason01}. These trends hold out to further distances as evidenced by multi-star companion surveys \cite{Raghavan10, Tokovinin14}, with additional binary star targets available to HabEx and LUVOIR. Many of these multi-star systems feature circumstellar dynamically stable habitable zones in which planets, if formed, would remain on stable long-term orbits. As a result, there is significant science that would be enabled by direct imaging of multi-star systems. 

The challenge due to the off-axis companion is caused by the off-axis leakage introduced in the region of interest being surveyed around the on-axis (target) star. When imaging with a telescope, there is a minimal amount of off-axis leakage introduced by diffraction from the telescope pupil. Additionally, aberrations at optical surface due to polishing of the primary and secondary mirrors as well as additional downstream imaging optics will introduce  off-axis leakage in the form of a speckle-field. In particular, these off-axis speckles are generated by higher-frequency components of the optical aberrations corresponding to the separation to the binary companion. Thus, intensity of the off-axis speckles is driven by:
\begin{itemize}
\item the brightness ratio between the target star and its off-axis companion
\item the angular separation between the multi-star components
\item the harmonic components of the power spectral density (PSD) across the region of interest
\end{itemize}
This off-axis leakage will result in a non-uniform contrast floor that will be difficult to remove via post-processing alone as the non-uniformity is caused by speckles.

\begin{figure}[h!]
\centering
\subfloat[]{
	\centering
	\label{fig:leakage-wfirstPupPSF}
	\includegraphics[width  = 0.45\columnwidth, trim = 1in 2.75in 0.45in 2.75in]{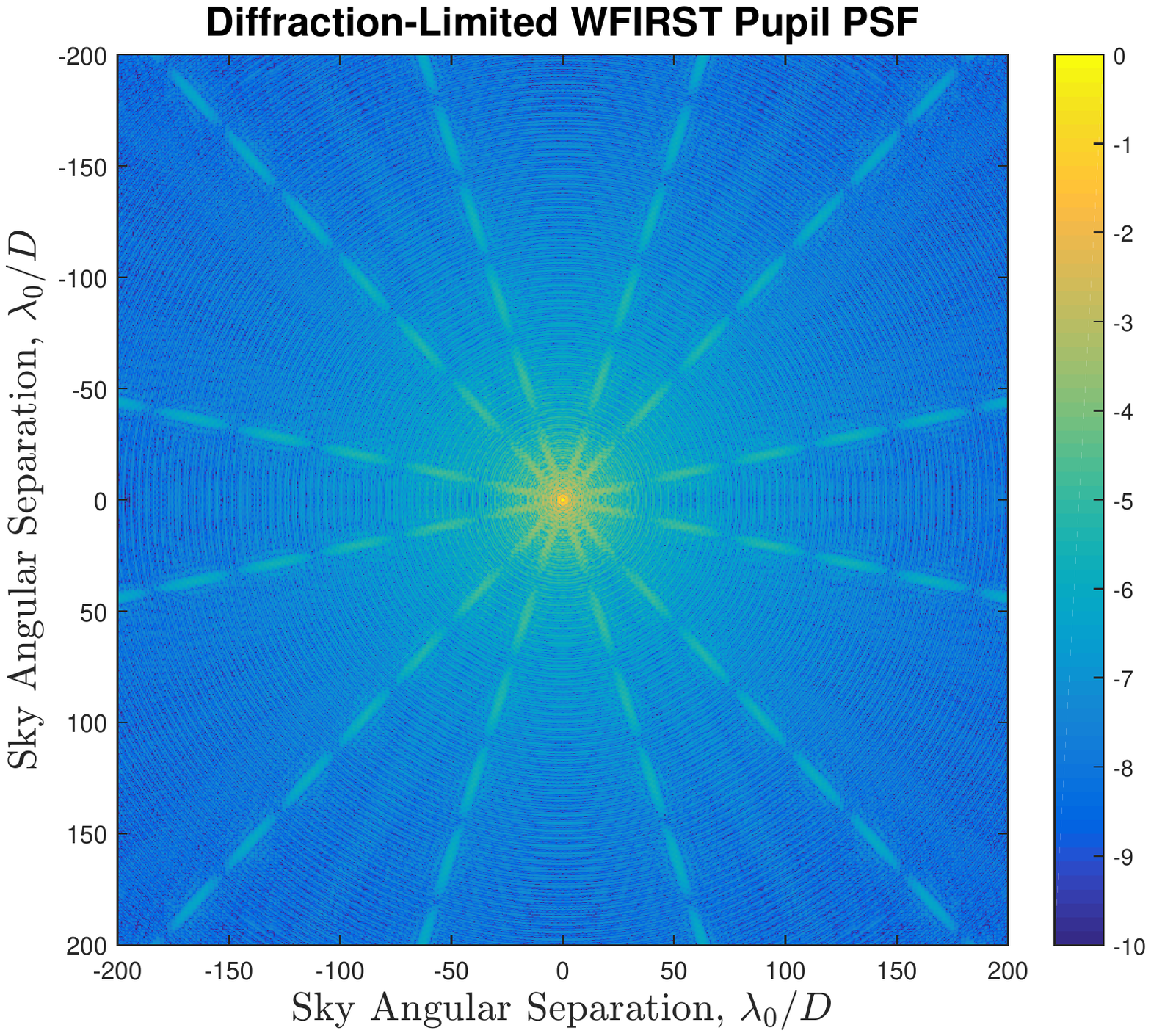}
}
\subfloat[]{
	\centering
	\label{fig:leakage-wfirstAbbPSF}
	\includegraphics[width  = 0.45\columnwidth, trim = 1in 2.75in 0.45in 2.75in]{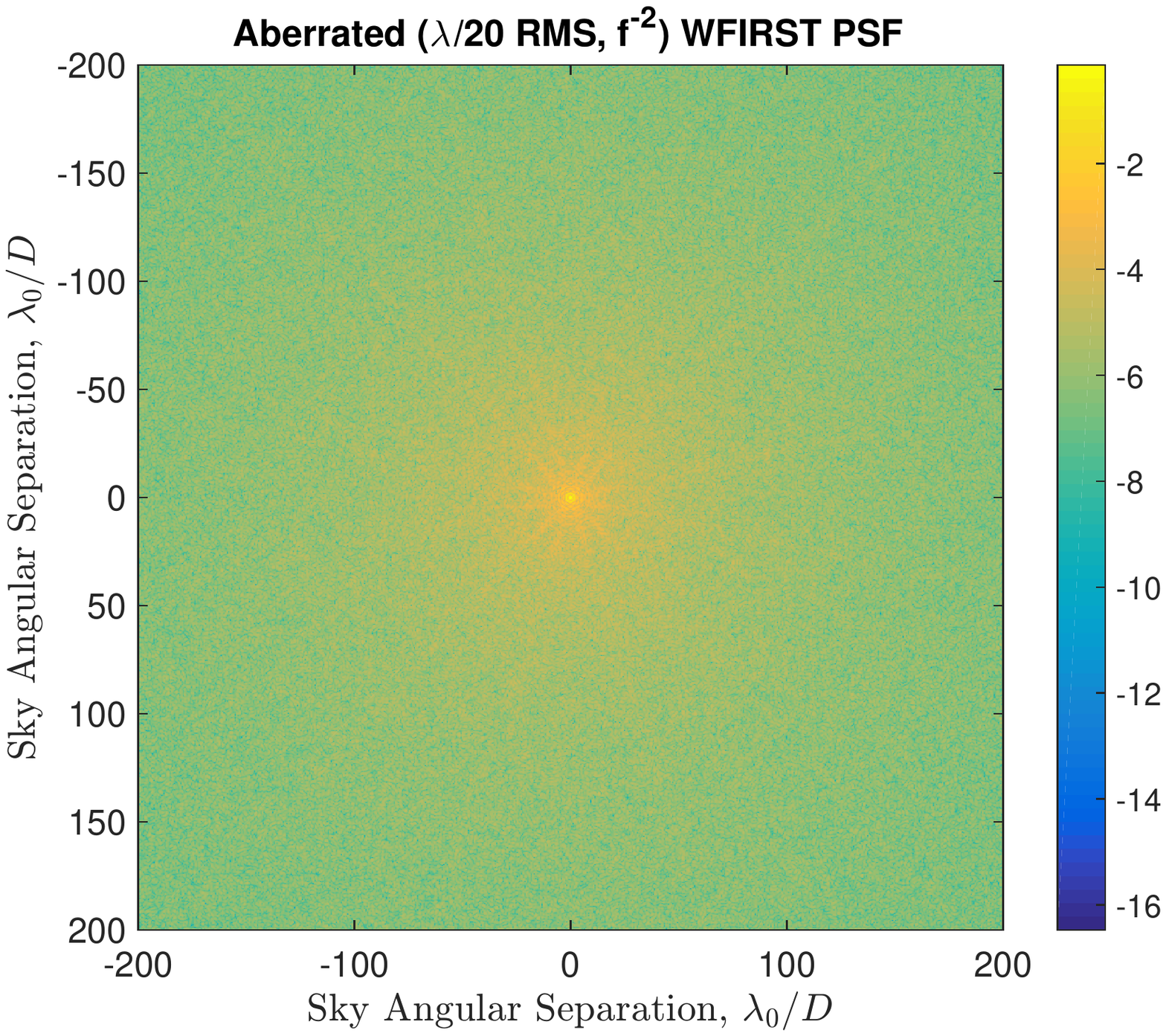}
}\\
\subfloat[]{
	\centering
	\label{fig:leakage-wfirstPup}
	\includegraphics[width  = 0.45\columnwidth, trim = 1in 2.75in 0.45in 2.75in]{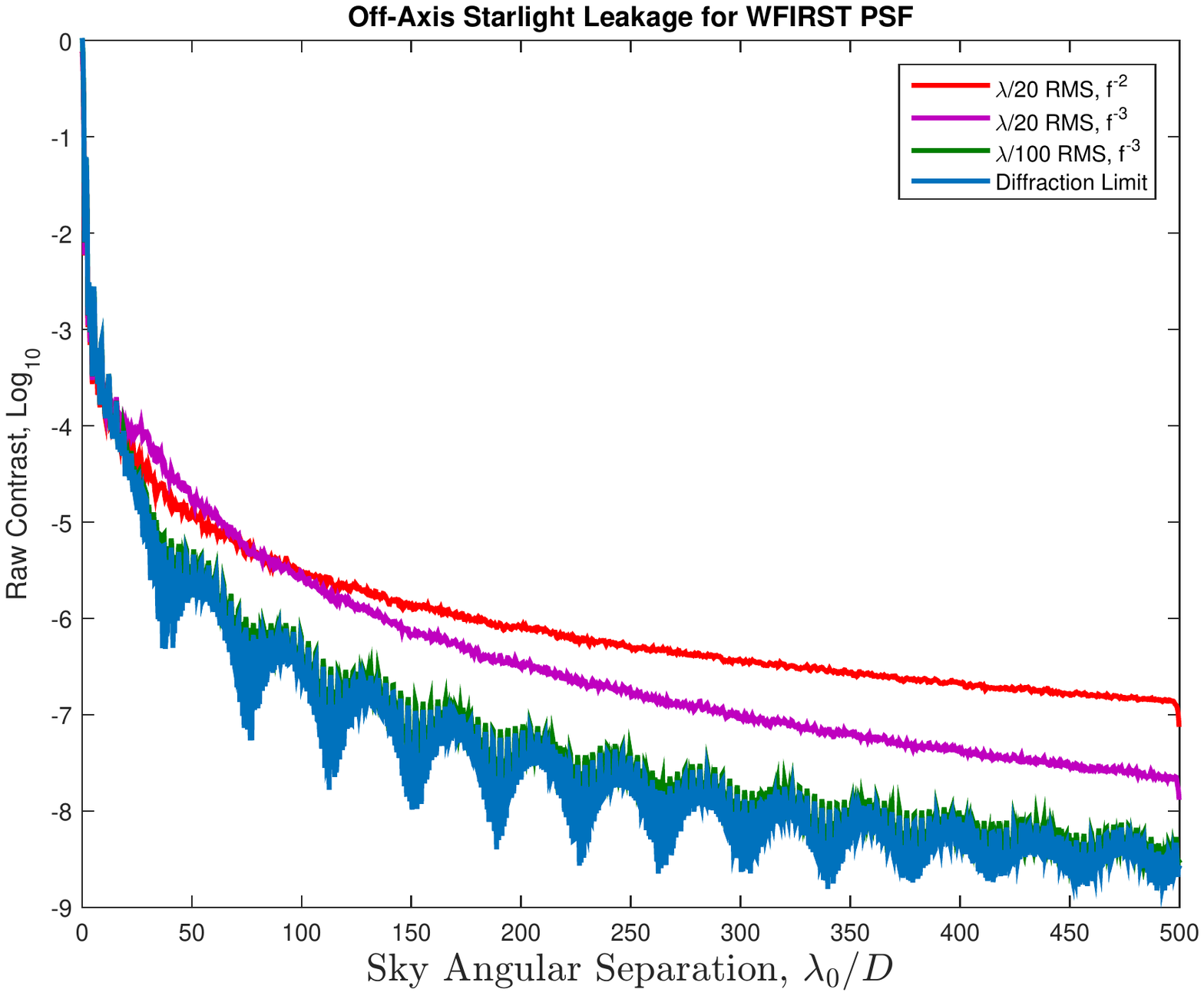}
}
\subfloat[]{
	\centering
	\label{fig:leakage-wfirstAbb}
	\includegraphics[width  = 0.45\columnwidth, trim = 1in 2.75in 0.45in 2.75in]{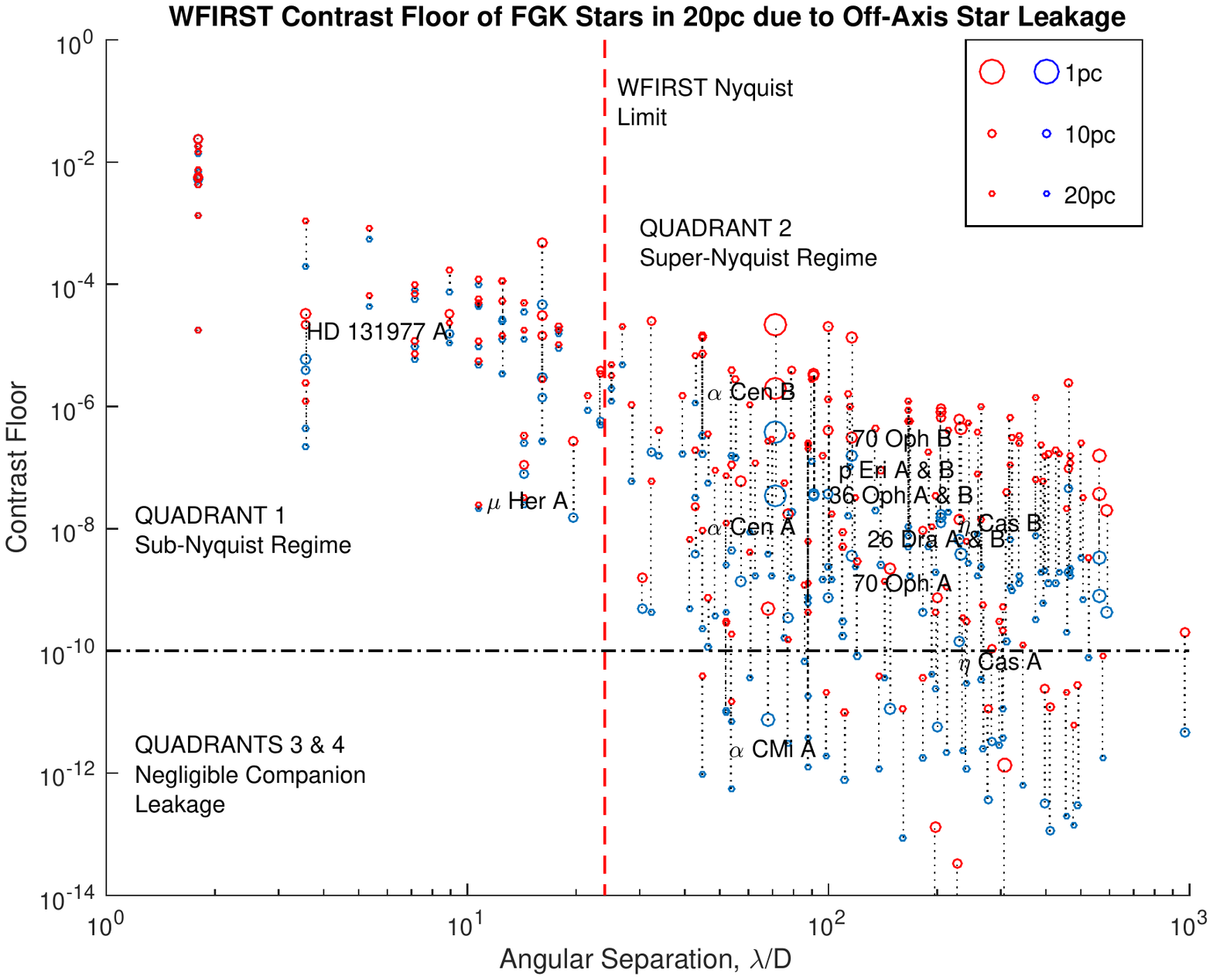}
}
\caption[Companion Leakage]
{\label{fig:leakage}  \subref{fig:leakage-wfirstPupPSF}  Diffraction-limited WFIRST Pupil PSF. \subref{fig:leakage-wfirstAbbPSF} Aberrated WFIRST PSF features speckles dominating at wider off-axis separations. \subref{fig:leakage-wfirstPup} WFIRST PSFs for different abberrated scenarios. Shown are curves corresponding to the diffraction-limited PSF, aberrations with $\lambda/100$ and $\lambda/20$ RMS for a PSD featuring a $1/f^{3}$ power spectral envelope, and $\lambda/20$ RMS for a PSD featuring a $1/f^{2}$  power spectral envelope. \subref{fig:leakage-wfirstAbb} Computed contrast floor due to the off-axis leakage from the binary companion for diffraction-limited WFIRST pupil (shown in blue) and for $\lambda/20$ RMS pure phase aberrations with a power spectral envelope following a $1/f^{2}$ power law (shown in red). Dotted lines indicate the contrast floor difference between these two scenarios for individual target stars.}
\end{figure}

To assess the impact of the off-axis leakage for a variety of targets of interest, we consider a simplified model of the WFIRST telescope consisting of a telescope pupil plane and an imaged focal plane without field stops. The results of this analysis are shown in Figure \ref{fig:leakage}. First, we consider a diffraction-limited PSF due to the WFIRST pupil (which contains spider obstructions and the secondary obscuration) in Figure \ref{fig:leakage-wfirstPupPSF}. This represents the diffraction leakage contribution which could be mitigated by usage of an appropriate coronagraph that controls the shape of the off-axis PSF. The corresponding azimuthal cross-section is depicted by the blue curve in Figure \ref{fig:leakage-wfirstPup}. The diffraction-limited PSF will be dominated by off-axis leakage due to speckles as the optical aberrations are introduced across the telescope. There is a high-degree of variance on the off-axis leakage intensity depending on the particular shape and total contribution of the aberration PSD. A conservative example of an aberrated WFIRST PSF constructed using $\lambda/20$ RMS aberrations with a PSF following a spectral envelope featuring a $1/f^2$ power law is shown in Figure \ref{fig:leakage-wfirstAbbPSF}. The azimuthal cross-section corresponding to this PSF is depicted by the red curve in Figure \ref{fig:leakage-wfirstPup}. Also shown are azimuthal curves corresponding to different aberation PSD assumptions: $\lambda/20$ RMS and $\lambda/100$ RMS with  power laws following a $1/f^{3}$ spectral envelope. 

Finally, in Figure \ref{fig:leakage-wfirstAbb}, the contrast floor due to the off-axis companion for all FGK multi-star targets within 20 pc is computed. These are computed for the optimistic case of a diffraction-limited PSF (shown in blue) as well as for the conservative case of an aberrated PSF with $\lambda/20$ RMS and a $1/f^2$ PSD envelope bounding the off-axis leakage variability possible depending on the aberration PSD. A dotted line connects the same stars under the two different scenario. This computation assumes again a WFIRST pupil with a diameter D = 2.4 m, computed for $\lambda = 650$ nm. The horizontal black line shows the $10^{-10}$ raw contrast target for imaging a rocky Earth-like planet. Between 140-180 stars out of 260 with known companions are located above the black-line --  for these stars the companion leakage must be suppressed. The vertical red line shows the Nyquist-limit of the WFIRST DM with 40 stars having close (sub-Nyquist) companion separations.

\begin{figure}[b!]
\centering
\subfloat[]{
	\centering
	\label{fig:alphaCen-historical}
	\includegraphics[width  = 0.4\columnwidth, trim = 0in 0in 0in 0in]{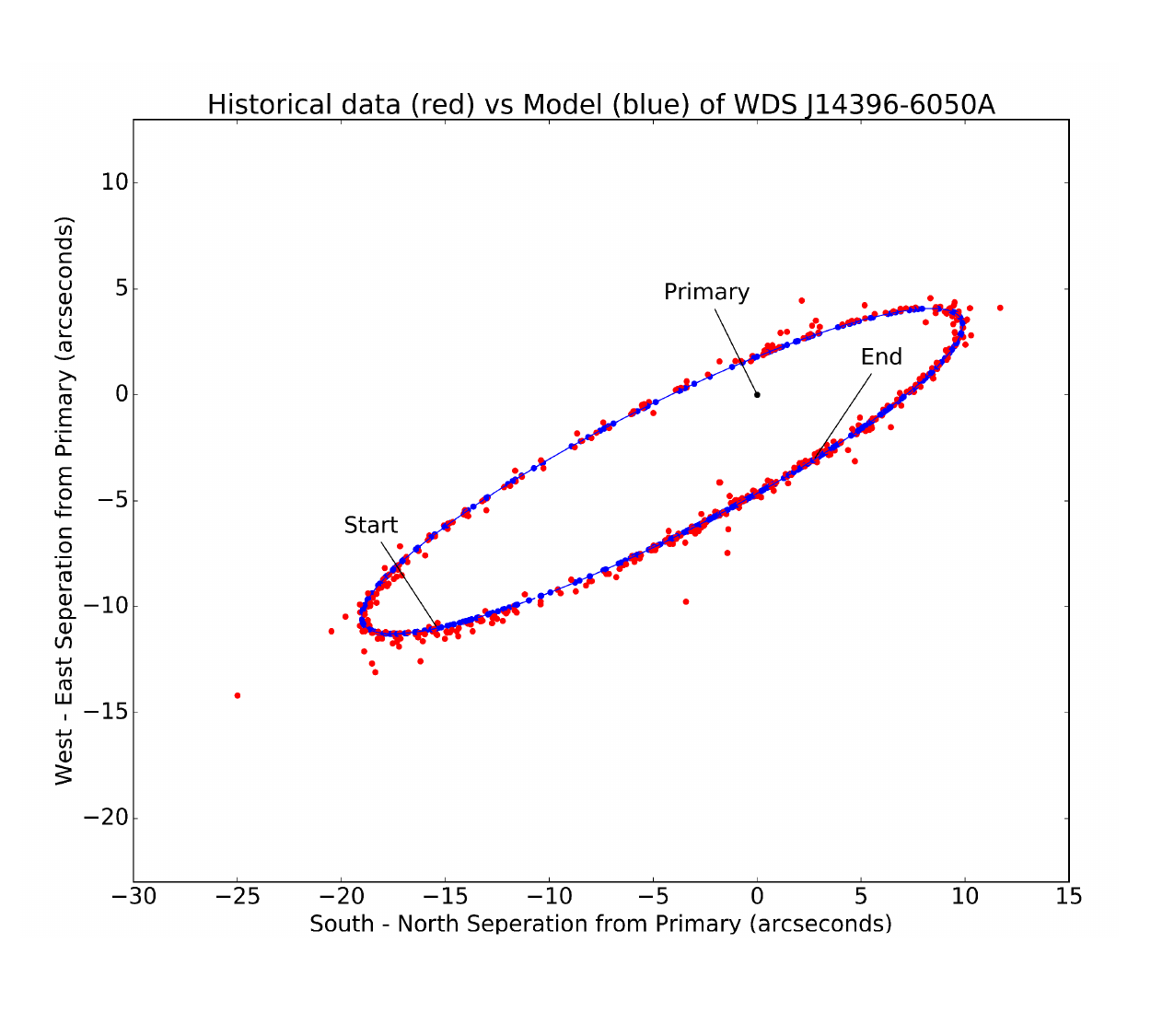}
}
\subfloat[]{
	\centering
	\label{fig:alphaCen-leak}
	\includegraphics[width  = 0.4\columnwidth, trim = 0in 0in 0in 0in]{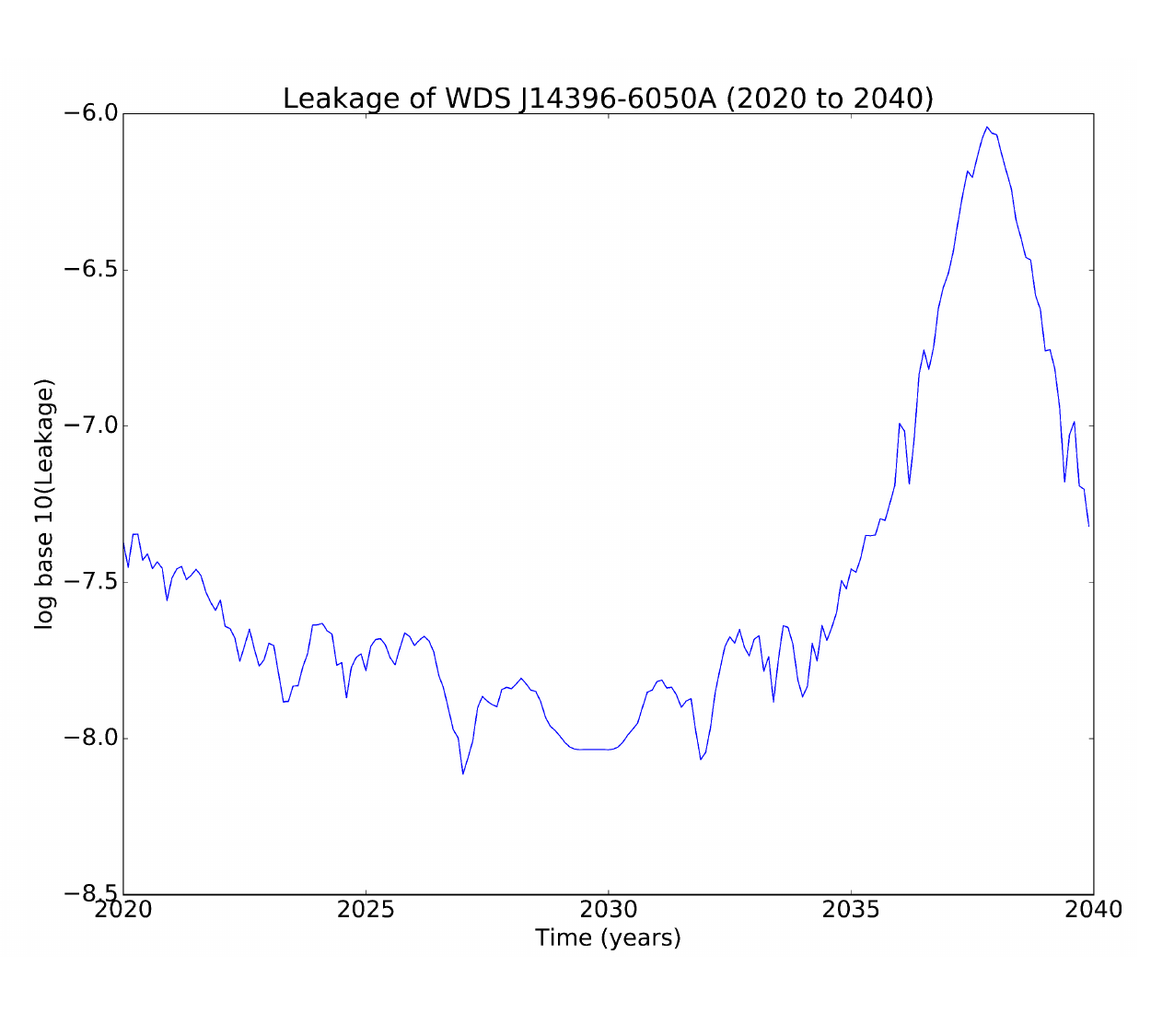}
}
\caption[Companion Leakage]
{\label{fig:alphaCen} \subref{fig:alphaCen-historical} Historical separation of Alpha Centauri companions shown as dotted lines and compared with ephemerides model \subref{fig:alphaCen-leak} Optimistic (near diffraction-limited) contrast floor for Alpha Centauri A star based on ephemerides-computed separations.}
\end{figure}

For example, the Alpha Centauri, located only 1.3 pc away and the nearest star system to the Sun, is a notable example of a binary star system -- Alpha Centauri A has a G2 spectral type while Alpha Centauri B has a K1 spectral type. The Alpha Centauri stars feature a brightness fraction of approximately 3, which results in a contrast-floor due to off-axis leakage 10x times higher when observing Alpha Centauri B than Alpha Centauri A. The stars are shown at approximately 4 arcsec separation (using the latest, 2016 entry, from WDS) which corresponds to 70 $\lambda/D$ for the WFIRST scenario described earlier. However, based on ORB6 orbital information the ephemerides of Alpha Centauri A and B can be computed and the separation predicted between 2020-2040 which would include the possible WFIRST observation window as shown in Figure \ref{fig:alphaCen}. During this period, the orbital separation between the companions would vary between 80-200 $\lambda/D$ with 180 $\lambda/D$ being the predicted separation in 2028. 

\section{Techniques for Controlling Off-Axis Leakage} \label{sect:leakage}

As discussed  in the previous section, a large fraction of multi-star targets requires suppression of the companion's leakage in order to image their circumstellar regions. There are two primary methods in which the off-axis star's leakage can be suppressed:
\begin{enumerate}
\item An off-axis starshade to block the binary companion
\item A wavefront control system to remove speckles from the binary companion
\end{enumerate}

\subsection{Off-Axis Starshade} \label{sect:offAxisStarshade}

\begin{figure}[b!]
\centering
\subfloat[]{
	\centering
	\label{fig:offAxisStarshade-before}
	\includegraphics[width  = 0.4\columnwidth, trim = 0in 2.75in 0in 2.75in]{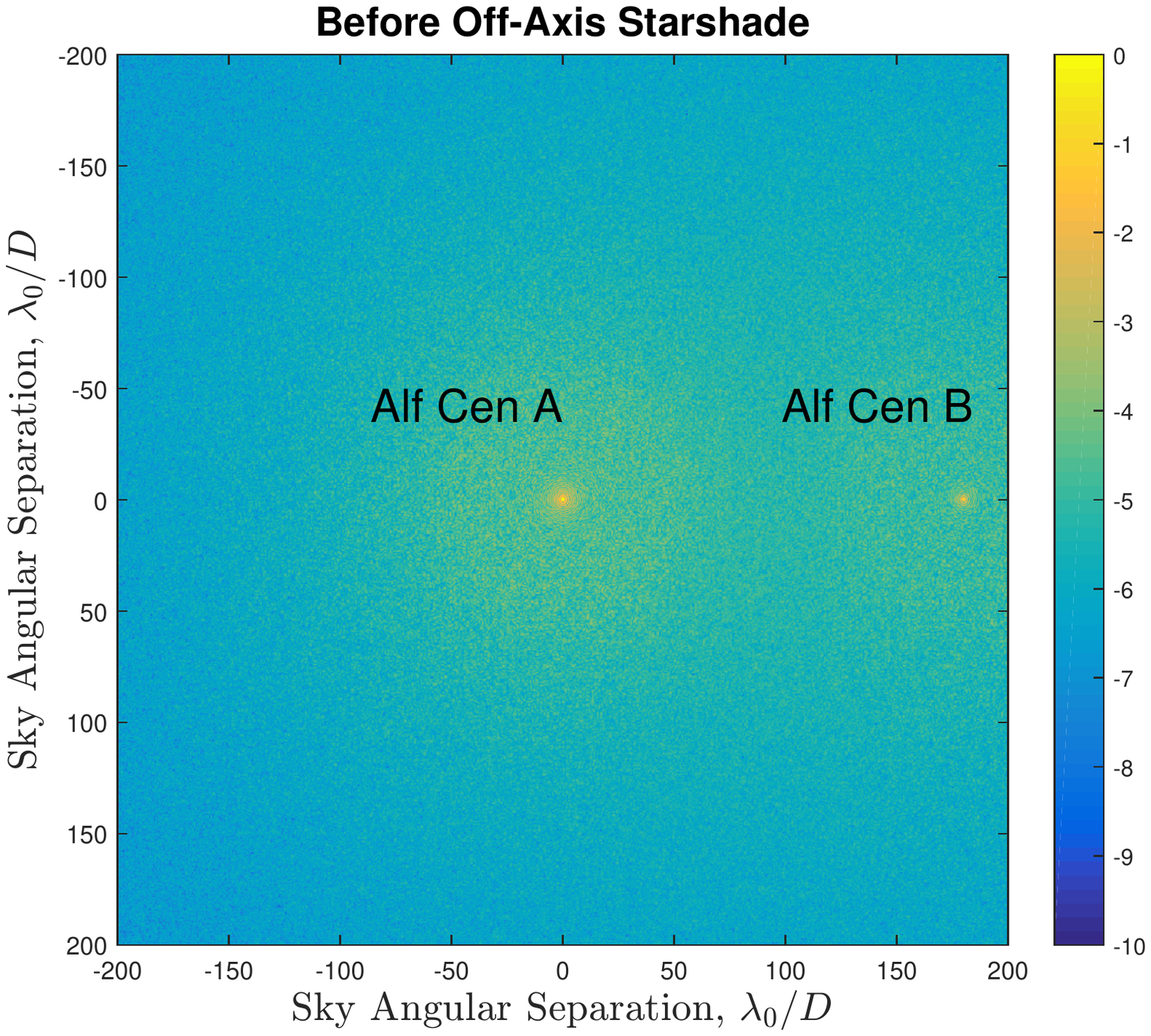}
}
\subfloat[]{
	\centering
	\label{fig:offAxisStarshade-after}
	\includegraphics[width  = 0.4\columnwidth, trim = 0in 2.75in 0in 2.75in]{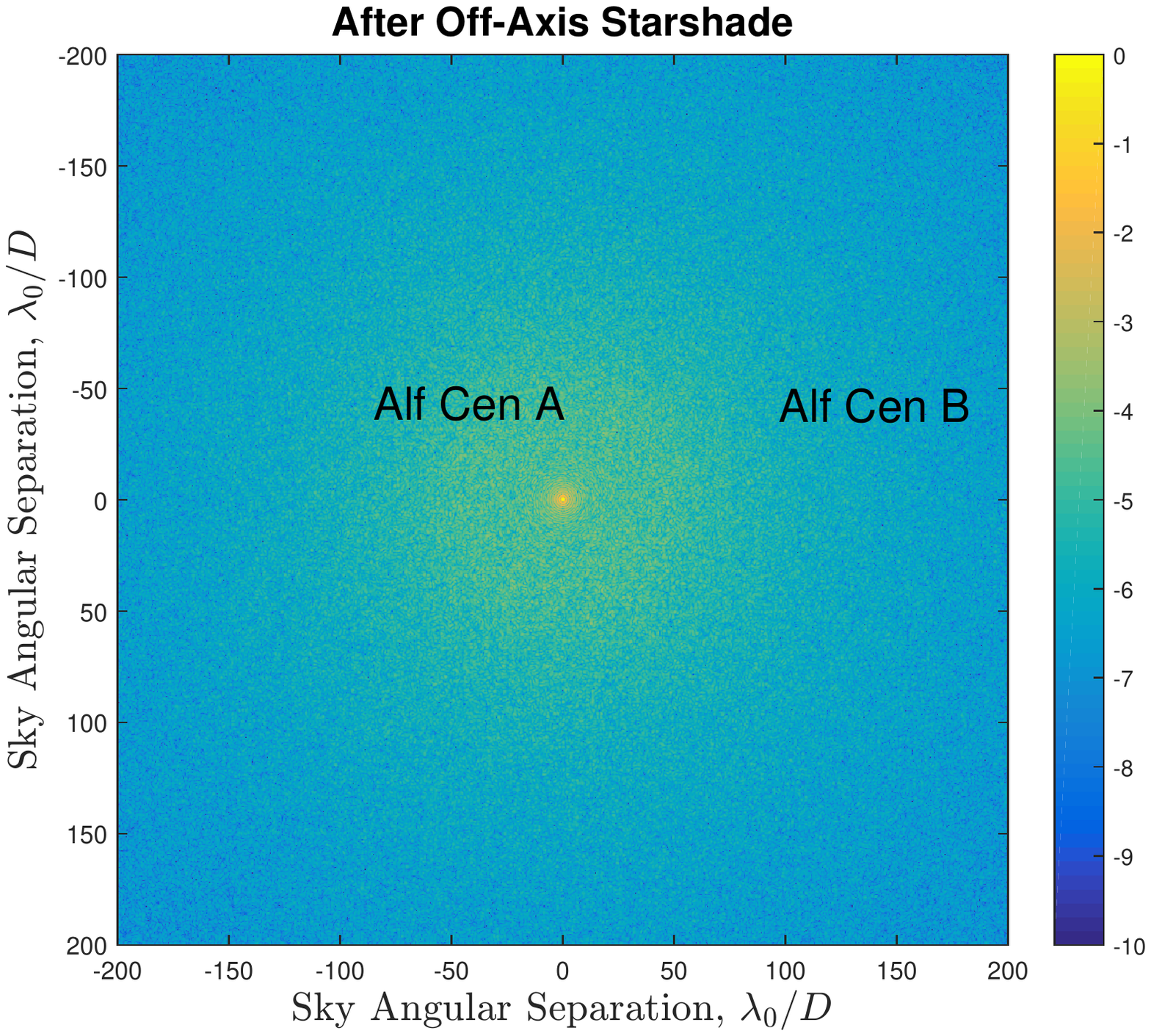}
}
\caption[Companion Leakage]
{\label{fig:offAxisStarshade} Usage of an off-axis starshade blocker \subref{fig:offAxisStarshade-before} Alpha Centauri A \& B before the off-axis starshade \subref{fig:offAxisStarshade-after} Alpha Centauri A \& B after the off-axis starshade.}
\end{figure}

A starshade can be operated off-axis in order to block the binary companion. Since the starshade suppresses starlight from the binary companion before it impinges on the telescope pupil, this method is not affected by optical aberrations across the telescope mirrors. Thus, the high-frequency components of the optical aberrations that cause the off-axis leakage described earlier have an insignificant effect when a starshade is used upstream of the telescope. 

The effect of the off-axis starshade is simulated in Figure \ref{fig:offAxisStarshade}. Both components of Alpha Centauri are clearly visible in Figure \ref{fig:offAxisStarshade-before}. When a starshade is in alignment and used to block Alpha Centauri B only the on-axis component of Alpha Centauri A is still visible. In this case, the WFIRST starshade provides 10 orders of magnitude suppression putting the off-axis PSF of Alpha Centauri B under the off-axis leakage of Alpha Centauri A. With the amount of suppression that a starshade can provide, Alpha Centauri B would not be visible even if Alpha Centauri A is diffraction limited. 

An off-axis starshade only needs to provide a few orders of magnitude suppression of the off-axis companion rather than the full 10 orders of magnitude for which it was designed. The amount of suppression necessary for a particular target can be inferred from the sample targets shown in Figure \ref{fig:leakage-wfirstAbb}. For close-in and bright companions the starshade would need to operate closer to 8 orders of magnitude suppression but for the vast majority of possible targets 4 orders of magnitude suppression would be sufficient. As a result, an off-axis starshade has reduced tolerances compared to operation on-axis in terms of both alignment and starshade edge errors. Additionally, it is possible that a smaller, reduced tolerance special-purpose off-axis starshade could be designed and flown to block the off-axis companion. This option, however, would still require launch and formation flight. 

\begin{figure}[t!]
\centering
\subfloat[]{
	\centering
	\label{fig:offAxisWC-beforeZoomOut}
	\includegraphics[width  = 0.4\columnwidth, trim = 1.25in 3.35in 1.25in 2.75in]{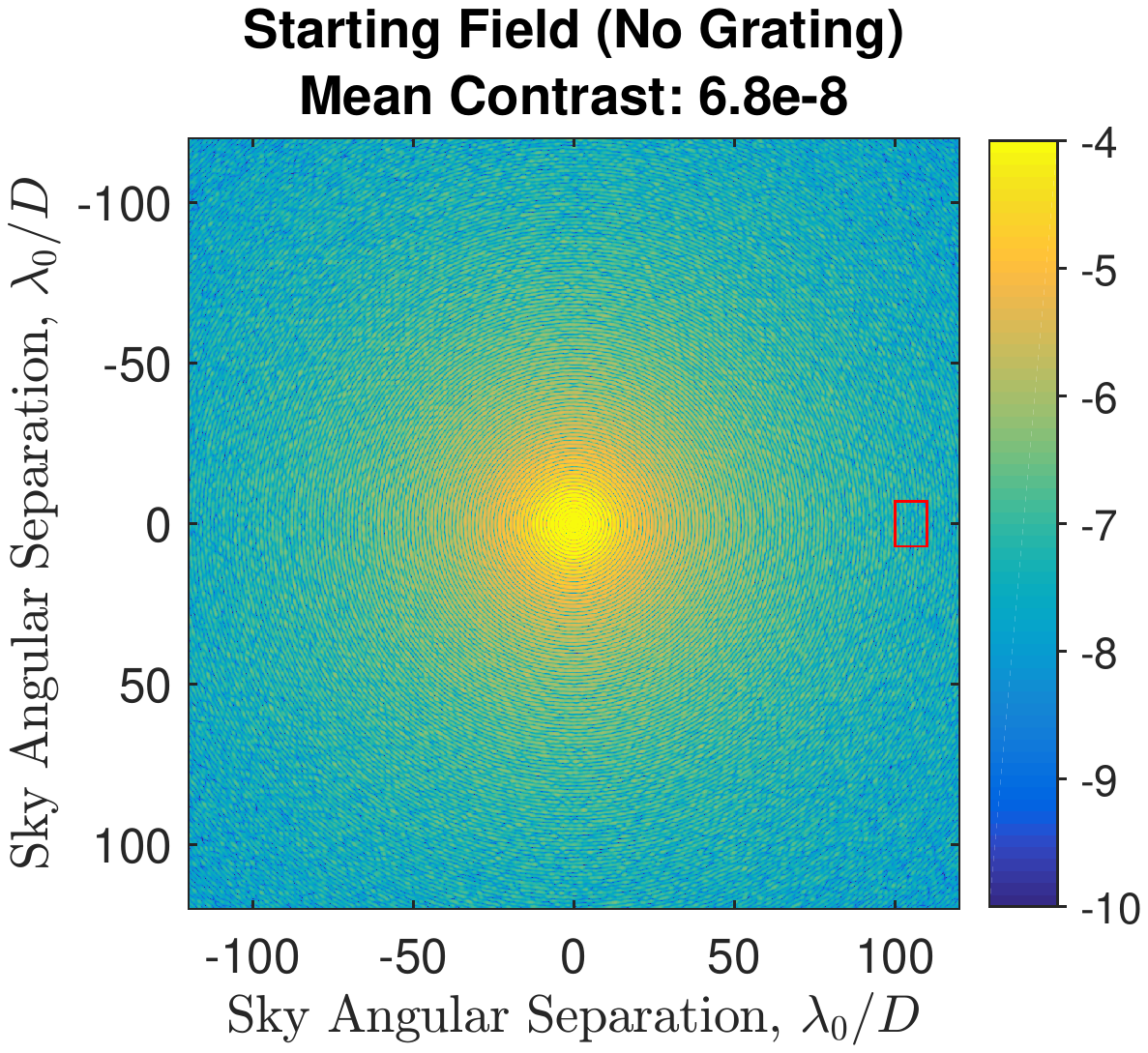}
}
\subfloat[]{
	\centering
	\label{fig:offAxisWC-beforeZoomIn}
	\includegraphics[width  = 0.4\columnwidth, trim = 1.25in 3.35in 1.25in 2.75in]{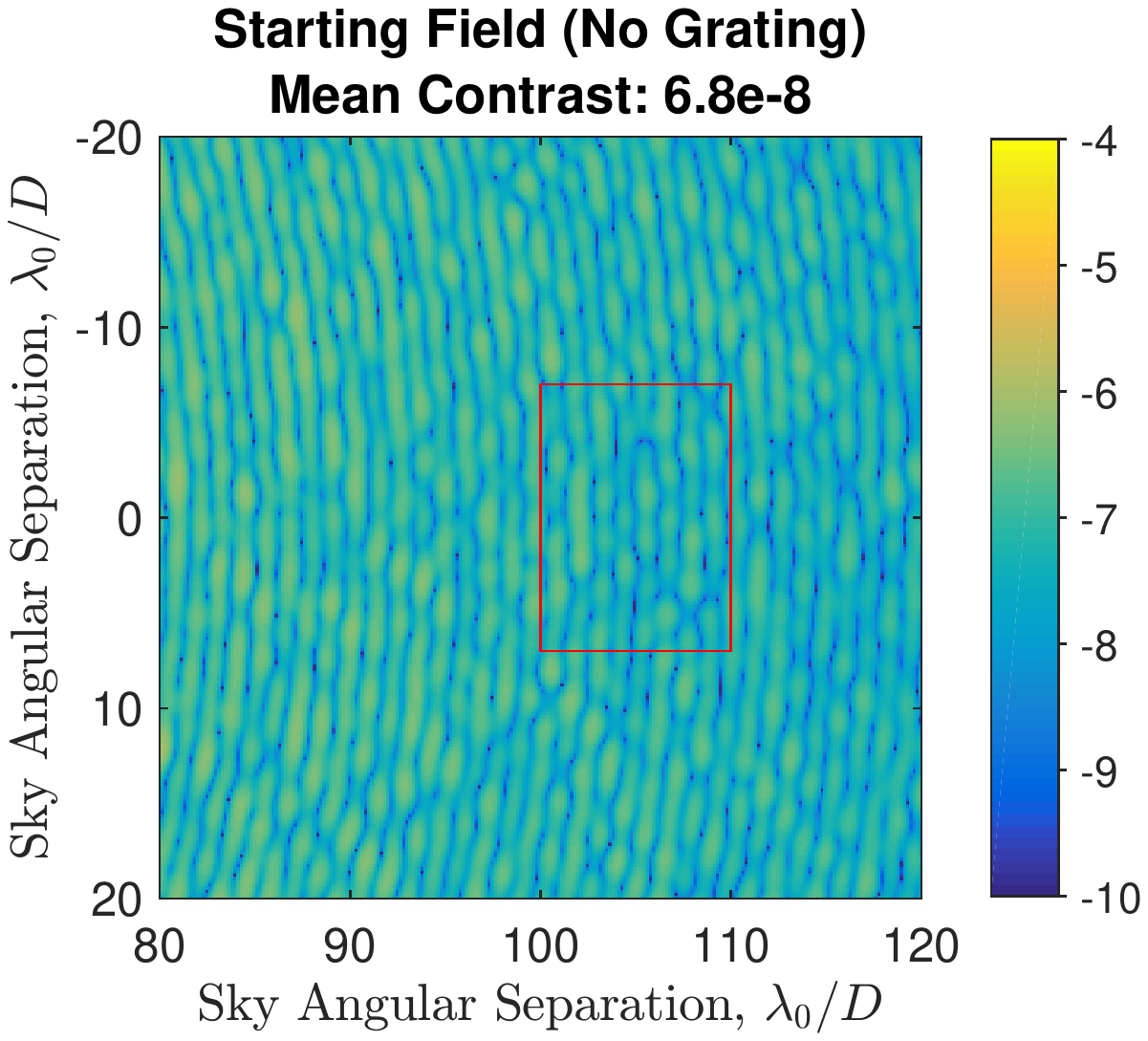}
}\\
\subfloat[]{
	\centering
	\label{fig:offAxisWC-afterZoomOut}
	\includegraphics[width  = 0.4\columnwidth, trim = 0in 0in 0in 0in]{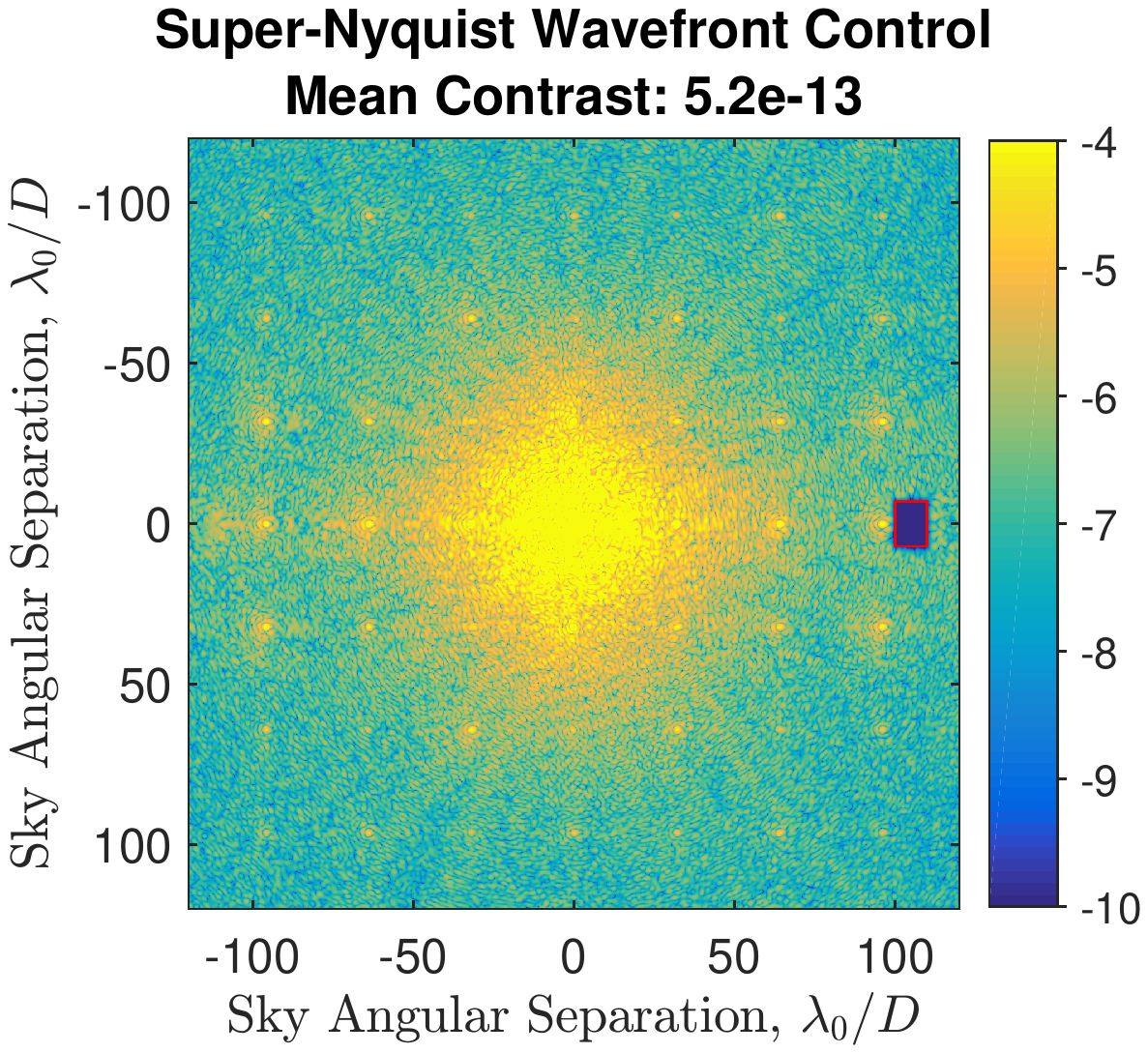}
}
\subfloat[]{
	\centering
	\label{fig:offAxisWC-afterZoomIn}
	\includegraphics[width  = 0.4\columnwidth, trim = 0in 0in 0in 0in]{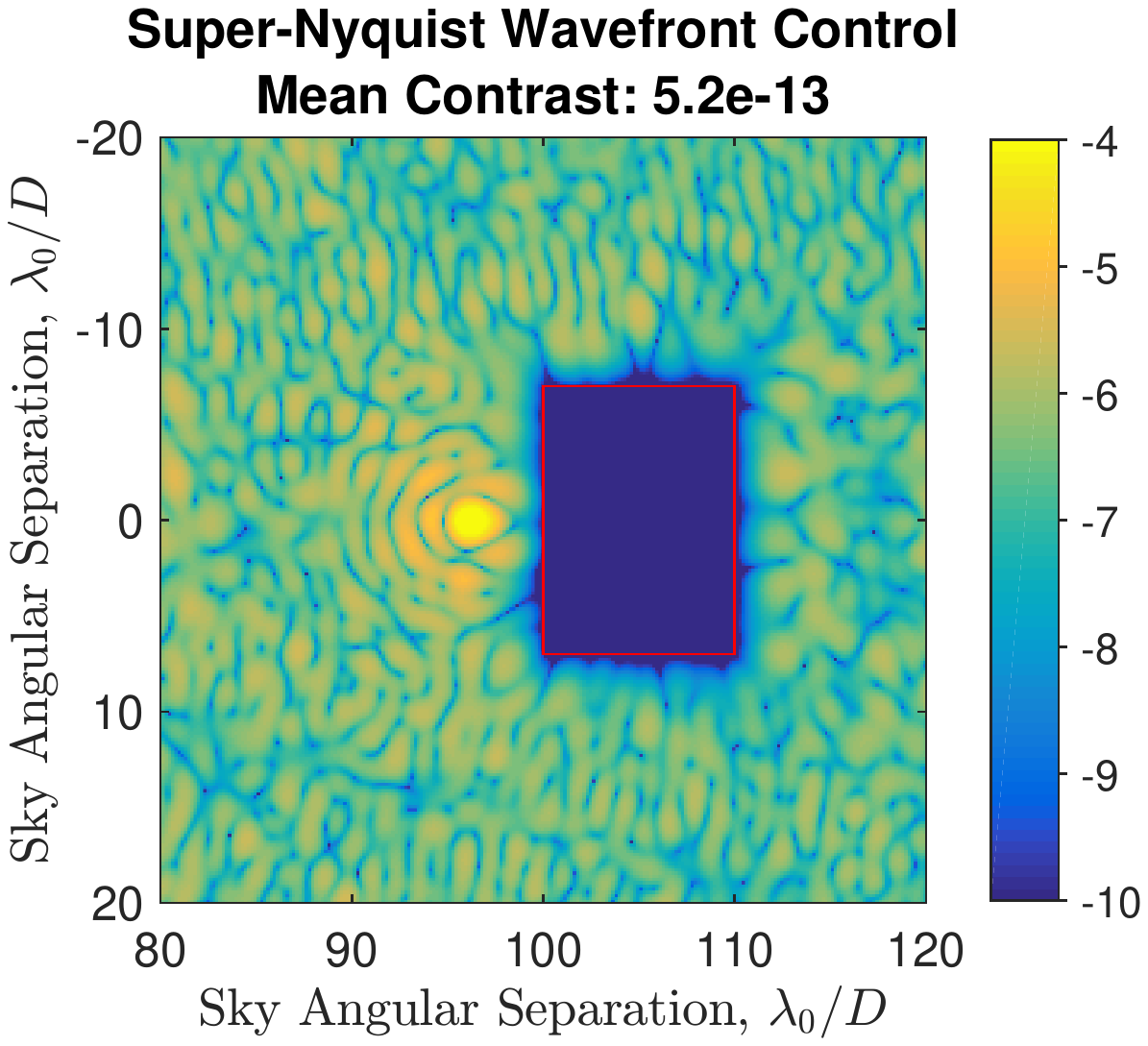}
}
\caption[Companion Leakage]
{\label{fig:offAxisWC} Usage of SNWC for off-axis leakage removal \subref{fig:offAxisWC-beforeZoomOut} Zoom-out of speckle field before SNWC and without the diffraction grating, \subref{fig:offAxisWC-beforeZoomIn} Zoom-in showing the control region before SNWC and without the diffraction grating, \subref{fig:offAxisWC-afterZoomOut} Zoom-out of speckle field after SNWC,  \subref{fig:offAxisWC-afterZoomIn} Zoom-in showing speckle field after SNWC}
\end{figure}

\subsection{Wavefront Control}

A wavefront control system using a DM can also be used to eliminate off-axis speckles from the binary companion after these are formed by the optical surfaces inside the telescope. If a coronagraph instrument is available on the telescope, only the wavefront control system is necessary to remove the off-axis speckles. 

An important consideration is the angular separation of the off-axis star. As apparent from Figure \ref{fig:leakage-wfirstAbb}, a small fraction of stars feature separations within the Nyquist limit of the DM. For these stars, the wavefront control system can be used to eliminate speckles as would be done normally for an on-axis star. We note here that it is possible to simultaneously and independently use the wavefront control system to eliminate speckles due to the on-axis star using a wavefront control technique referred to as Multi-Star Wavefront Control (MSWC). MSWC\cite{Sirbu17} uses non-redundant modes on the DM to enable multi-star imaging. Recent laboratory results demonstrating MSWC performance are presented in this conference \cite{Belikov17}.

For the large fraction of stars for which the angular separation is beyond the Nyquist limit of the DM a diffraction grating is necessary to enable Super-Nyquist Wavefront Control (SNWC) and control speckles from the off-axis star at wider angular separations. Operation of SNWC (introduced and discussed in more detail elsewhere \cite{Thomas15}) is summarized in Figure \ref{fig:offAxisWC}. In Figure \ref{fig:offAxisWC-beforeZoomOut}, the initial monochromatic speckle field from a star is shown -- the region of interest in which off-axis speckles are to be removed is marked, with an initial contrast level of $6.8 \times 10^{-8}$. These are pin-speck aberrations at a similar level to the Airy rings of a circular pupil as can be seen from the zoomed-in image in Figure \ref{fig:offAxisWC-beforeZoomIn}.   

Addition of a diffraction grating creates a set of regular diffraction orders that depend on the spacing of the structure on the grating. The effect is to create a set of PSF replicas across the field of view. In monochromatic light the replica PSF structure and peaks are clearly visible across the image in Figure \ref{fig:offAxisWC-afterZoomOut}. The PSF replicas allow coherent modulation using the DM regions within the Nyquist limit of the control diffraction order. Thus, wavefront control algorithms such as Electric Field Conjugation (EFC) or speckle nulling can now be employed to eliminate speckles at super-Nyquist separations. Such a resulting dark-hole is shown in Figure \ref{fig:offAxisWC-afterZoomIn}. Although the diffraction grating worsens the initial contrast, the DM is able to eliminate both the additional diffracted light from the grating as well as the original off-axis leakage. In this example, the final contrast across region of interest is $5.2 \times 10^{-13}$, which is nearly 5 orders of magnitude below the mean raw contrast of the original speckle field. 

\section{Simulated WFIRST Scenarios} \label{sect:wfirst}

There are a number of options available to enable multi-star imaging scenarios for using starshades. We consider three separate scenarios for direct imaging of Alpha Centauri as follows:
\begin{enumerate}
\item Dual starshades: an on-axis starshade in conjunction with on off-axis starshade
\item On-axis WFIRST Coronagraph Instrment (CGI) in conjunction with an off-axis starshade
\item On-axis starshade with SNWC operating off-axis 
\end{enumerate}
These simulated scenarios will focus on the WFIRST configuration, but these same basic scenarios are possible with HabEx or LUVOIR. Additionally, we note that it is possible to use MSWC alone without a starshade to enable multi-star direct imaging. 

\begin{figure}[t!]
\centering
\subfloat[]{
	\centering
	\label{fig:dualStarshades-before}
	\includegraphics[width  = 0.45\columnwidth, trim = 0in 0in 0in 0in]{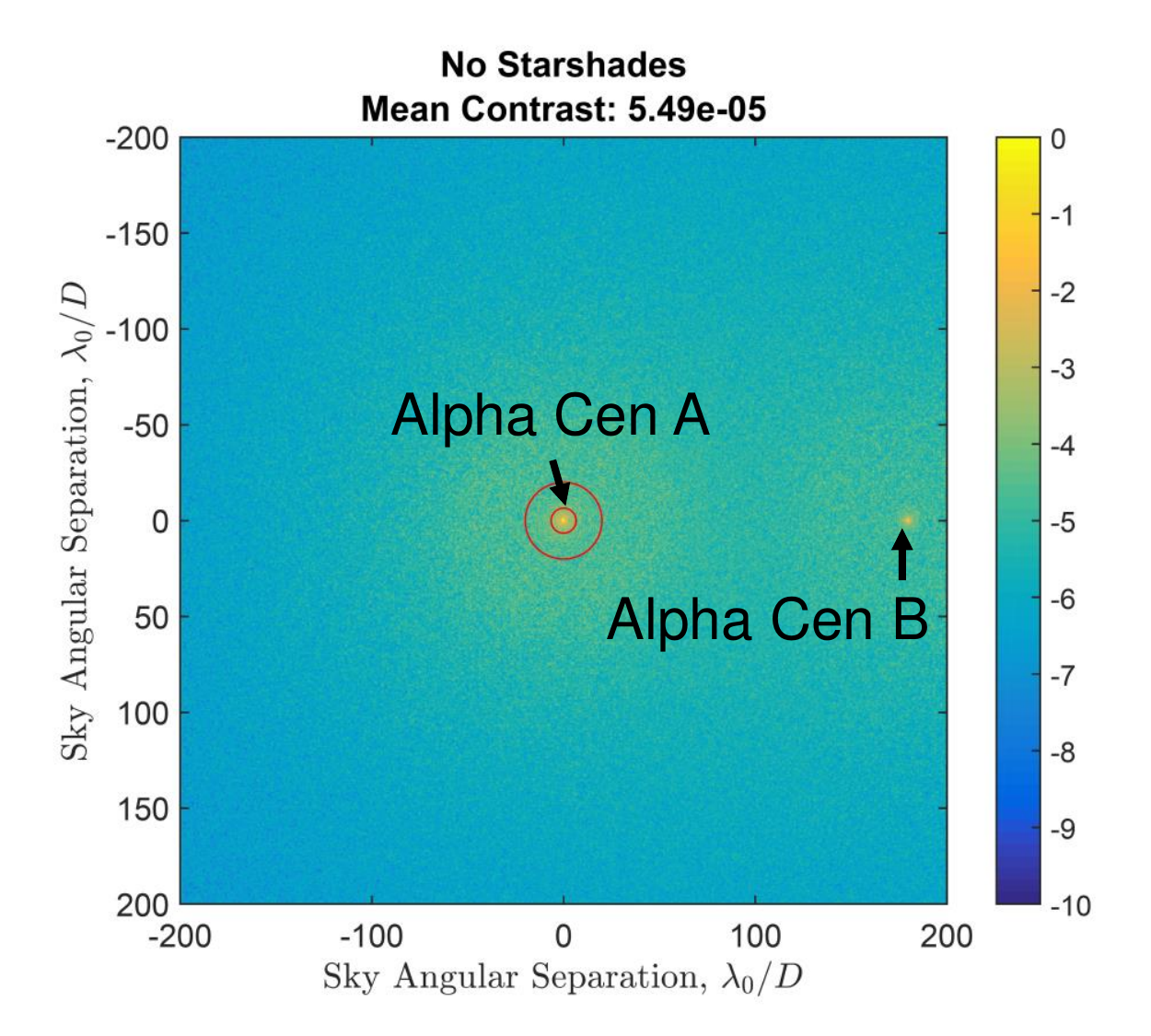}
}
\subfloat[]{
	\centering
	\label{fig:dualStarshades-after}
	\includegraphics[width  = 0.45\columnwidth, trim =0in 0in 0in 0in]{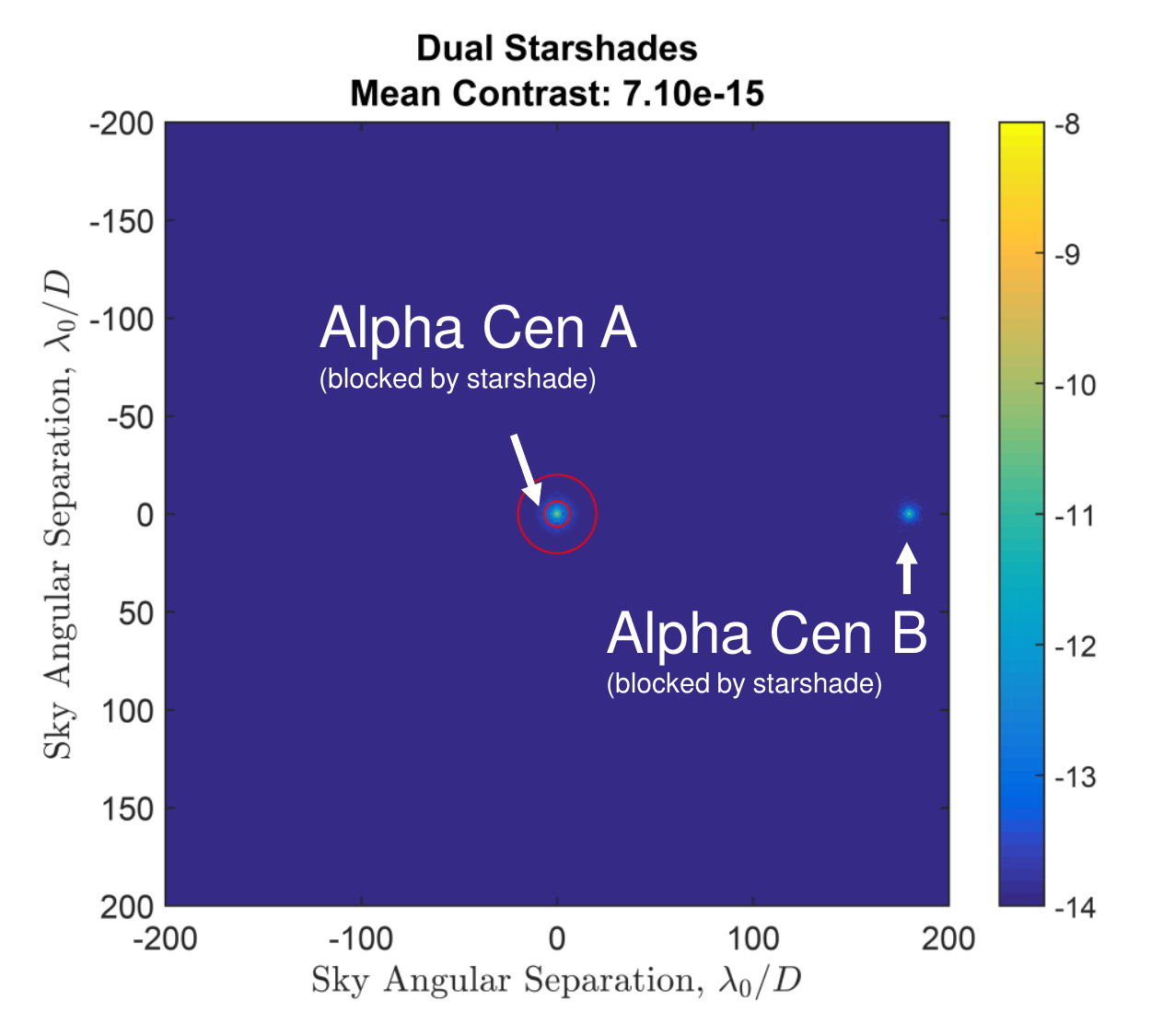}
}
\caption[Companion Leakage]
{\label{fig:dualStarshades} Demonstration of multi-star imaging with WFIRST using dual starshades \subref{fig:dualStarshades-before} Before dual-starshades \subref{fig:dualStarshades-after} After dual-starshades}
\end{figure}

\subsection{Dual Starshades}

The most basic option to enable multi-star imaging with starshades is to use dual starshades. Thus, one starshade operates on-axis blocking the main star. A second starshade operates off-axis to block off-axis leakage as described in Section \ref{sect:offAxisStarshade}.

This scenario is simulated in Figure \ref{fig:dualStarshades}. To model this scenario, two identical starshades with 40-m diameter at 55,000 km separation and featuring near-ideal diffraction performance casts a 4-m shadow with ten orders of magnitude suppression across the 2.4 m WFIRST telescope pupil. The electric field is propagated using two-dimensional Fresnel propagation at 650 nm \cite{Sirbu16AO}. The Alpha Centauri stars featuring a 180 $\lambda/D$ angular separation are shown before the starshades are in position in Figure \ref{fig:dualStarshades-before} with contrast across the Alpha Centauri A habitable  zone limited by the on-axis star at approximately $5.5 \times 10^{-5}$ and off-axis leakage from Alpha Centauri B in combination with optical aberrations across the WFIRST telescope creating a contrast floor at the $10^{-7}$ level. Once the dual starshades are introduced, both stars are suppressed by nearly ten orders of magnitude resulting in a $10^{-15}$ contrast across the Alpha Centauri A region of interest. 

The advantage of the dual starshade method is that the entire habitable zone can be directly imaged (and exoplanets spectroscopically characterized) at high-contrast out to $20 \lambda/D$ in 20\% band. The achievable outer working angle and characterization bandwidth are driven by the WFIRST IFS specifications adjusted for maximum starshade accommodation \cite{Mandell17}. The disadvantage of a dual starshade operation is that it requires two starshades to be available; additionally, this option requires the second starshade to act as an off-axis blocker during observation and thus the second starshade cannot be slewed to the next target star while observing multi-star systems. Whereas the off-axis starshade alignment tolerances can be reduced, there is nonetheless a requirement to align the starshade off-axis while outside the nominal field of view of the science camera or the on-axis Low-Order Wavefront Sensor (LOWFS) sensor \cite{Bottom17}.

\begin{figure}[t!]
\centering
\subfloat[]{
	\centering
	\label{fig:cgiStarshade-before}
	\includegraphics[width  = 0.45\columnwidth, trim = 0in 0in 0in 0in]{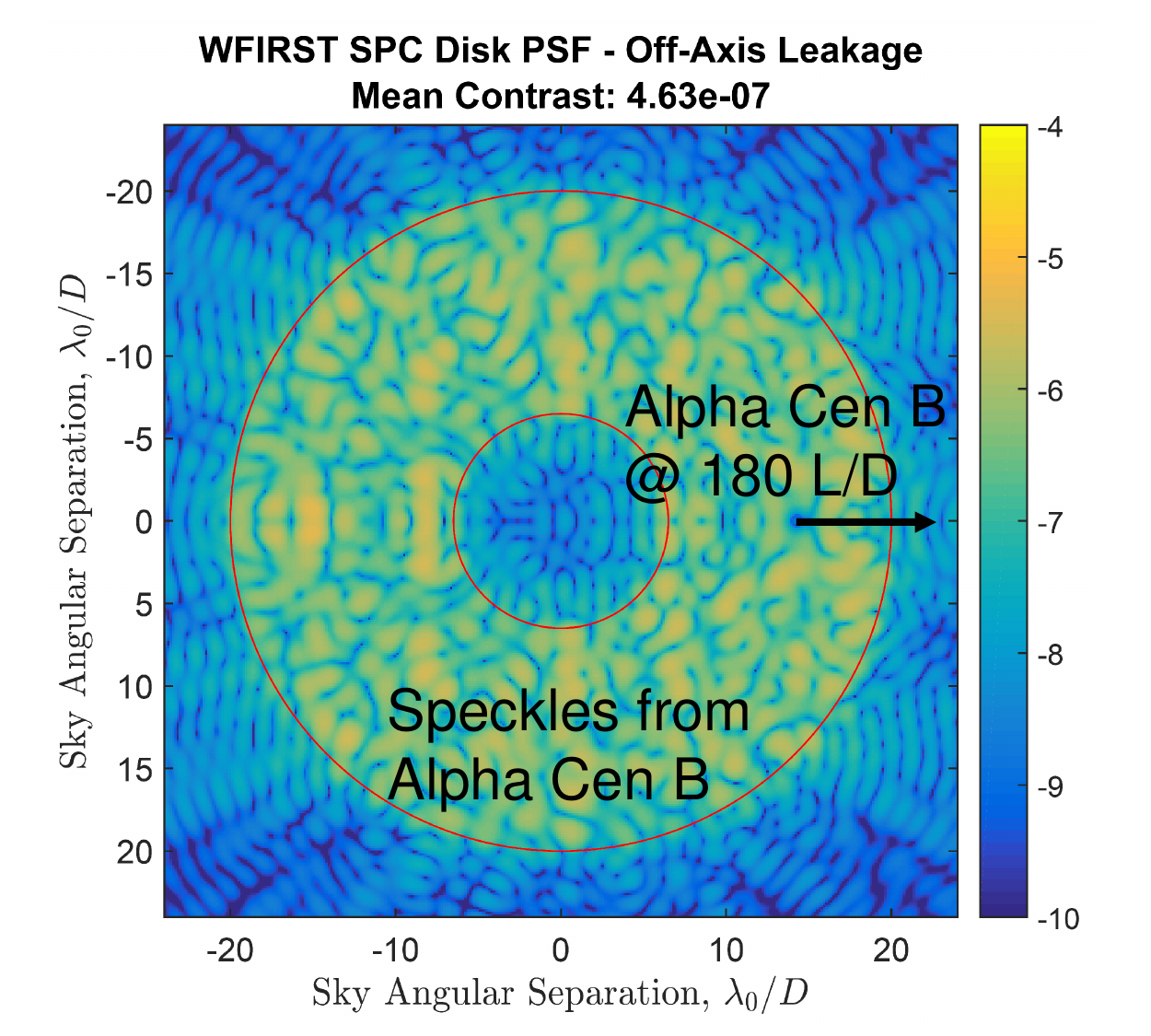}
}
\subfloat[]{
	\centering
	\label{fig:cgiStarshade-after}
	\includegraphics[width  = 0.45\columnwidth, trim =0in 0in 0in 0in]{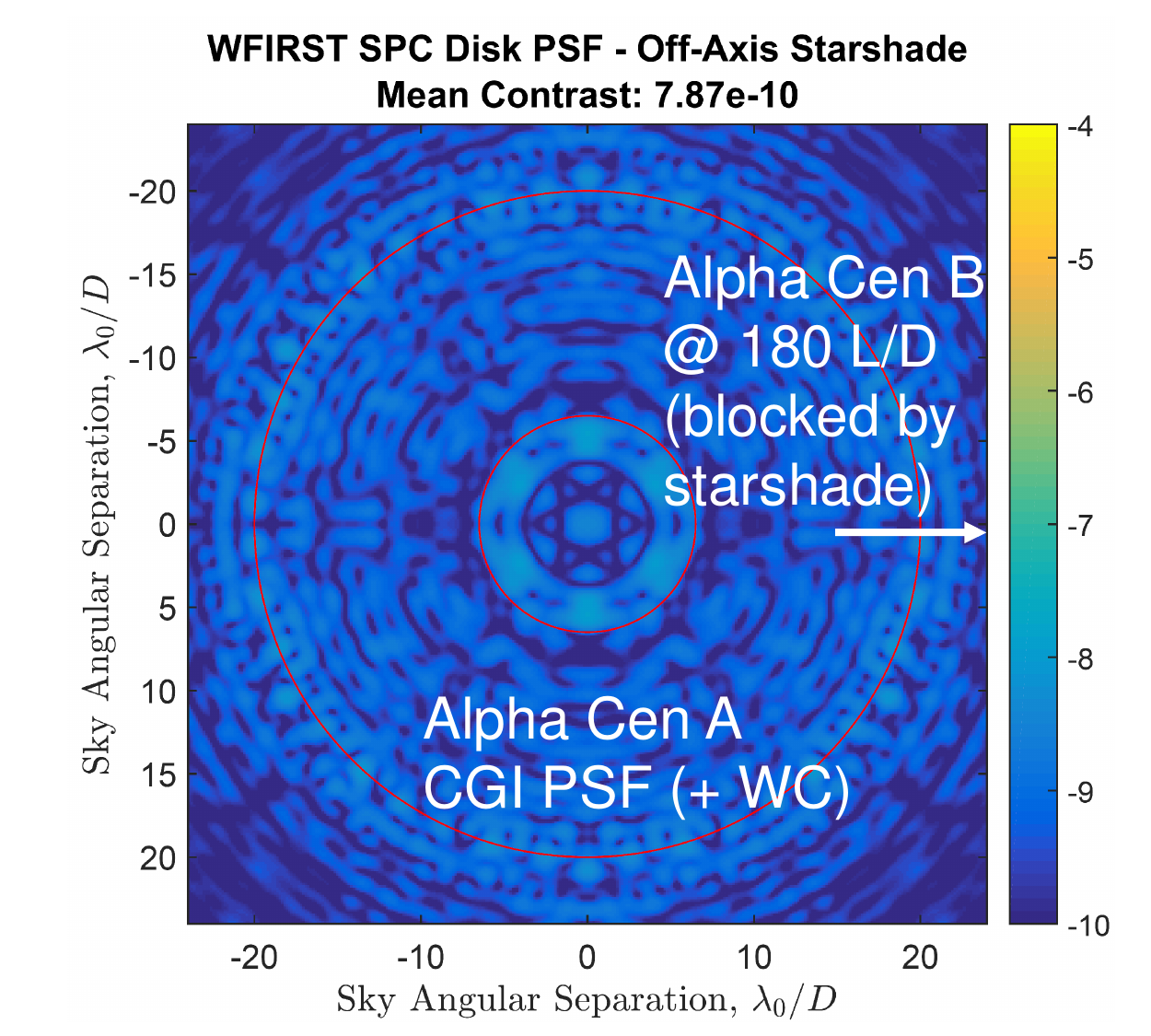}
}
\caption[Companion Leakage]
{\label{fig:cgiStarshade} Demonstration of multi-star imaging with WFIRST using the on-axis Shape Pupil Coronagraph (SPC) Disk Mask with an onff-axis Starshade \subref{fig:cgiStarshade-before} Before off-axis starshade (limited by off-axis speckles) \subref{fig:cgiStarshade-after} After off-axis starshade showing SPC Disk Mask PSF.}
\end{figure}

\subsection{On-Axis CGI with Off-Axis Starshade}

A second option is to use the WFIRST CGI in hybrid operation with a potential WFIRST starshade. The WFIRST CGI could be operated normally and the starshade used off-axis. This would effectively turn the multi-star imaging problem into a single-star imaging scenario as long as the off-axis starshade alignment is maintained.

The best CGI observation channel would be the Shaped Pupil Coronagraph (SPC) Disk Mask which has a nominal field of view from 6.5-20 $\lambda_0 / D$ and is coupled into the IFS for spectroscopic characterization \cite{Riggs17}. The SPC Disk Mask has a 360-degree field of view and would cover almost the complete habitable zone of Alpha Centauri A and the complete habitable zone of Alpha Centauri B. The WFIRST CGI would operate normally with the DM removing speckles from the on-axis target star with an expected raw contrast of about $10^{-9}$. This scenario is summarized in Figure \ref{fig:cgiStarshade}. Shown in Figure \ref{fig:cgiStarshade-before} is the incoherent WFIRST CGI Disk mask PSF (assumed to be after wavefront control with on-axis speckles) and the off-axis speckles due to Alpha Centauri B. The contrast is thus limited by the off-axis speckles from Alpha Centauri B at $4.63 \times 10^{-7}$. Once the off-axis starshade is aligned, the off-axis speckles from Alpha Centauri B are suppressed by ten orders of magnitude and are thus far below the CGI SPC Disk Mask's PSF which is shown in Figure \ref{fig:cgiStarshade-after}.

The disadvantage of this method is that the full achievable raw contrast of a designed starshade operating with WFIRST is not achievable with contrast limited by the WFIRST pupil. Additionally, the WFIRST SPC Disk mask is designed for operation at 10\% bandwidth requiring two sets of observations for characterization at 20\% bandwidth. Finally, the starshade would have to be aligned off-axis outside the LOWFS and science camera field of view. However, no operation or hardware changes would be necessary to the WFIRST CGI which would be able to operate as designed. Additionally, this option could potentially enable rocky planet imaging only around the Alpha Centauri stars (due to its proximity, the Alpha Centauri system could provide sufficient Signal-to-Noise after speckle removal in post-processing and enable rocky-planet planet detection), but further stars would be limited by the CGI raw contrast.

\begin{figure}[t!]
\centering
\subfloat[]{
	\centering
	\label{fig:starshadeSNWC-before}
	\includegraphics[width  = 0.45\columnwidth, trim = 0in 0in 0in 0in]{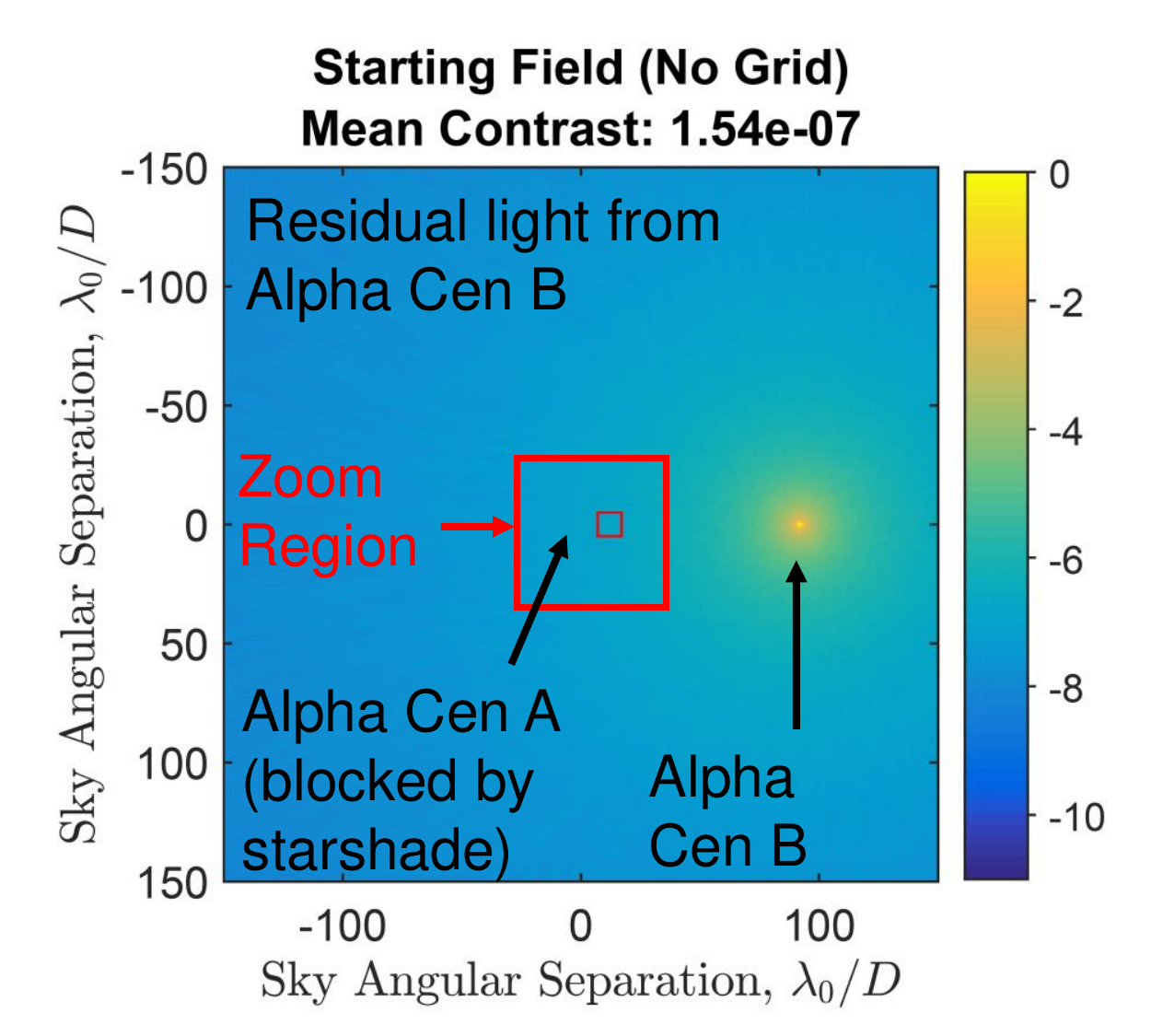}
}
\subfloat[]{
	\centering
	\label{fig:starshadeSNWC-after}
	\includegraphics[width  = 0.45\columnwidth, trim =0in 0in 0in 0in]{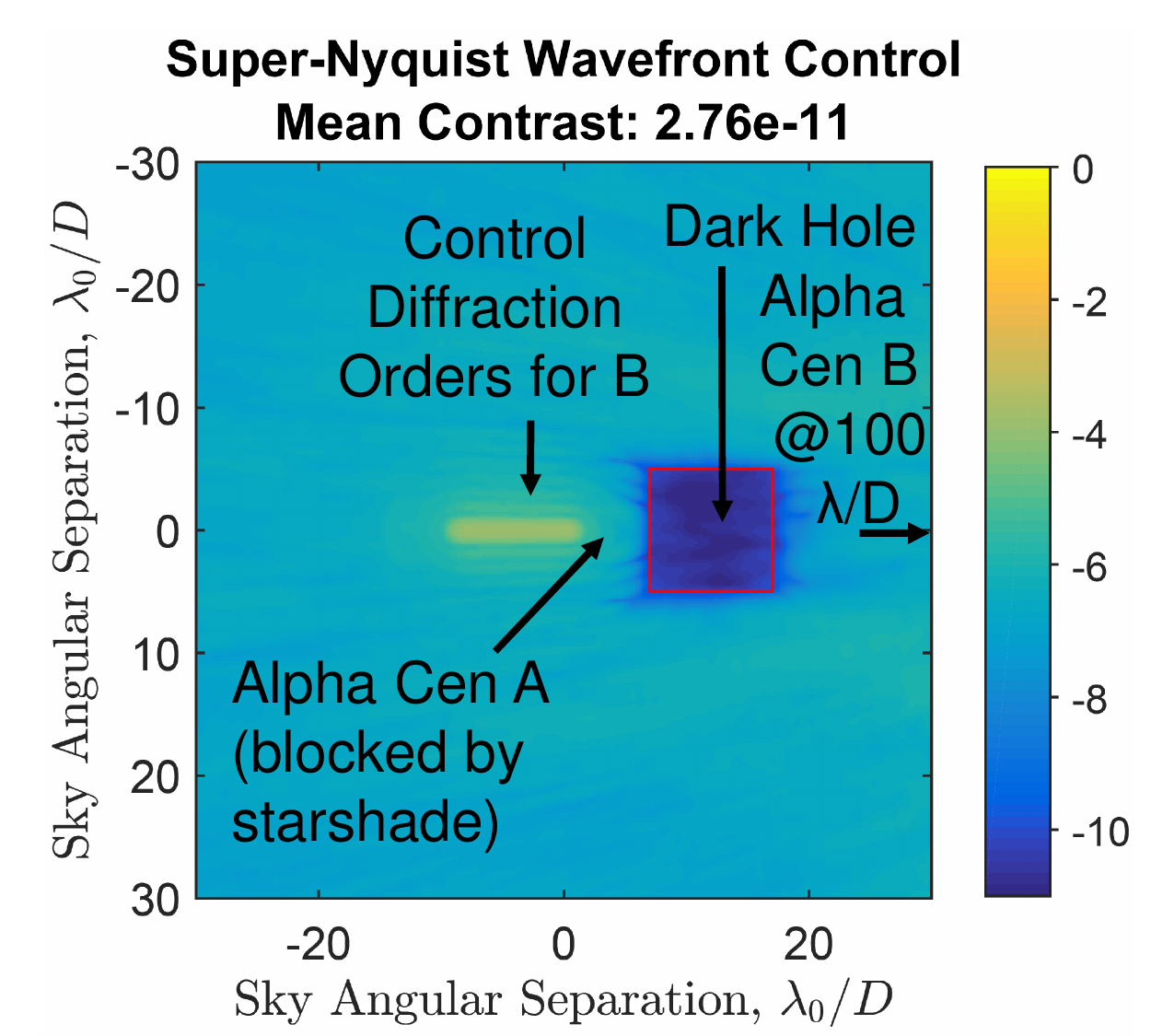}
}
\caption[Companion Leakage]
{\label{fig:starshadeSNWC} Demonstration of on-axis starshade with off-axis leakage controlled via SNWC \subref{fig:starshadeSNWC-before} Before off-axis SNWC (without diffraction grating) \subref{fig:starshadeSNWC-after} After SNWC in 10\% bandwidth.}
\end{figure}

\subsection{On-Axis Starshade with Super Nyquist Wavefront Control}

A final option would be to use a starshade on-axis and to remove the off-axis speckles using the wavefront control system available on the WFIRST CGI. In this scenario, a diffraction grating would have to be available upstream of any focal-plane masks blocking the field of view. This could be realized by inserting an amplitude diffraction grating mask consisting of a set of pupil-plane dots at the SPC filter location. Another option would be to take advantage of the set of quilting orders naturally occurring as part of the DM fabrication process. The CGI SPC channel would be used to enable IFS characterization without using a focal plane mask (since the on-axis star is blocked by the starshade).

This scenario is simulated in Figure \ref{fig:starshadeSNWC} with the Alpha Centauri stars separated by 100 $\lambda/D$. The starshade is deployed and blocking the on-axis star (Alpha Centauri A) in Figure \ref{fig:starshadeSNWC-before}. As a result, the field of view is dominated by residual light from Alpha Centauri B. In order to create a region of high-contrast around Alpha Centauri A then, it is necessary to eliminate speckles from Alpha Centauri B. In the indicated $10 \lambda/D \times 10 \lambda/D$ region of interest the mean contrast due to the off-axis leakage from Alpha Centauri B  is $1.54 \times 10^{7}$. The off-axis speckles from Alpha Centauri B are then removed using Super-Nyquist Wavefront Control (SNWC) as shown in Figure \ref{fig:starshadeSNWC-after}. The control diffraction order used to extend the controllability region is prominent. This is elongated due to the smearing at different wavelengths for a 10\% bandwidth and operating at relatively large off-axis separations (100 $\lambda/D$). After SNWC, the achieved contrast is $2.76 \times 10^{-11}$ contrast enabling circumstellar imaging around Alpha Centauri A in the generated dark hole. In this scenario, SNWC was able to suppress off-axis speckles by nearly four orders of magnitude.

\begin{figure}[t!]
\centering
\subfloat[]{
	\centering
	\label{fig:snwcObservation-discover-zoomOut}
	\includegraphics[width  = 0.45\columnwidth, trim = 0in 0in 0in 0in]{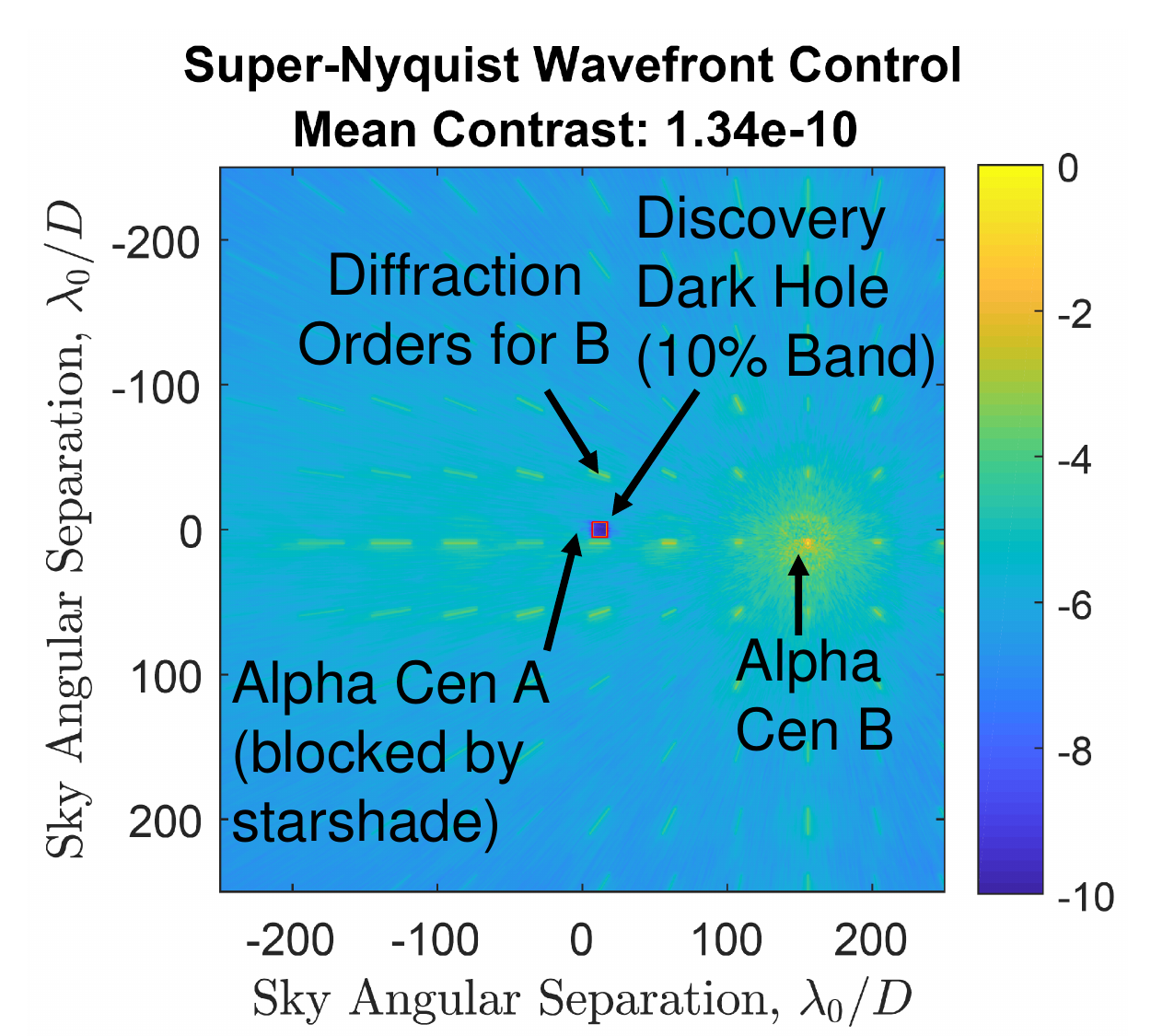}
}
\subfloat[]{
	\centering
	\label{fig:snwcObservation-discover-zoomIn}
	\includegraphics[width  = 0.45\columnwidth, trim =0in 0in 0in 0in]{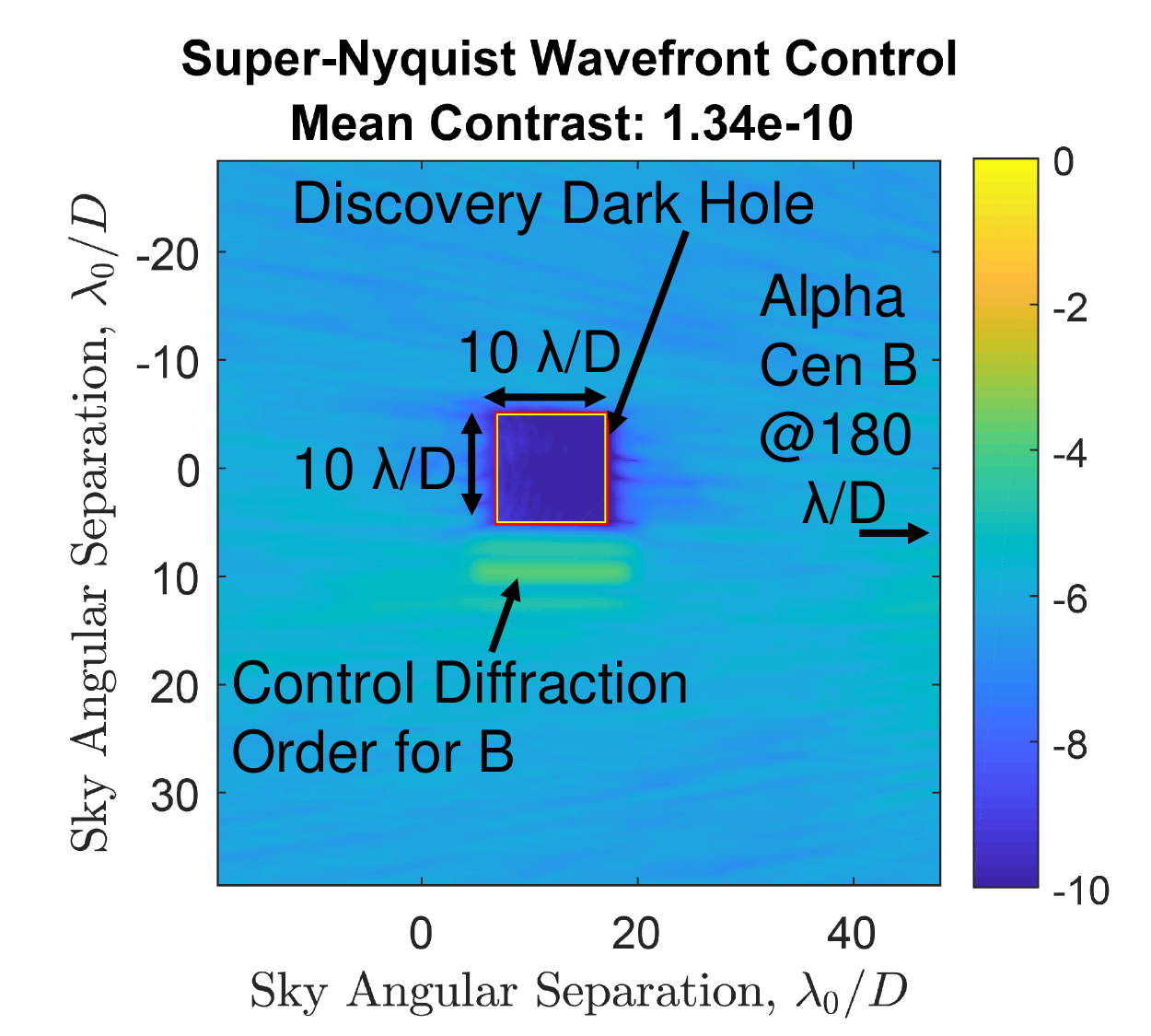}
}\\
\centering
\subfloat[]{
	\centering
	\label{fig:snwcObservation-characterization-zoomOut}
	\includegraphics[width  = 0.45\columnwidth, trim = 0in 0in 0in 0in]{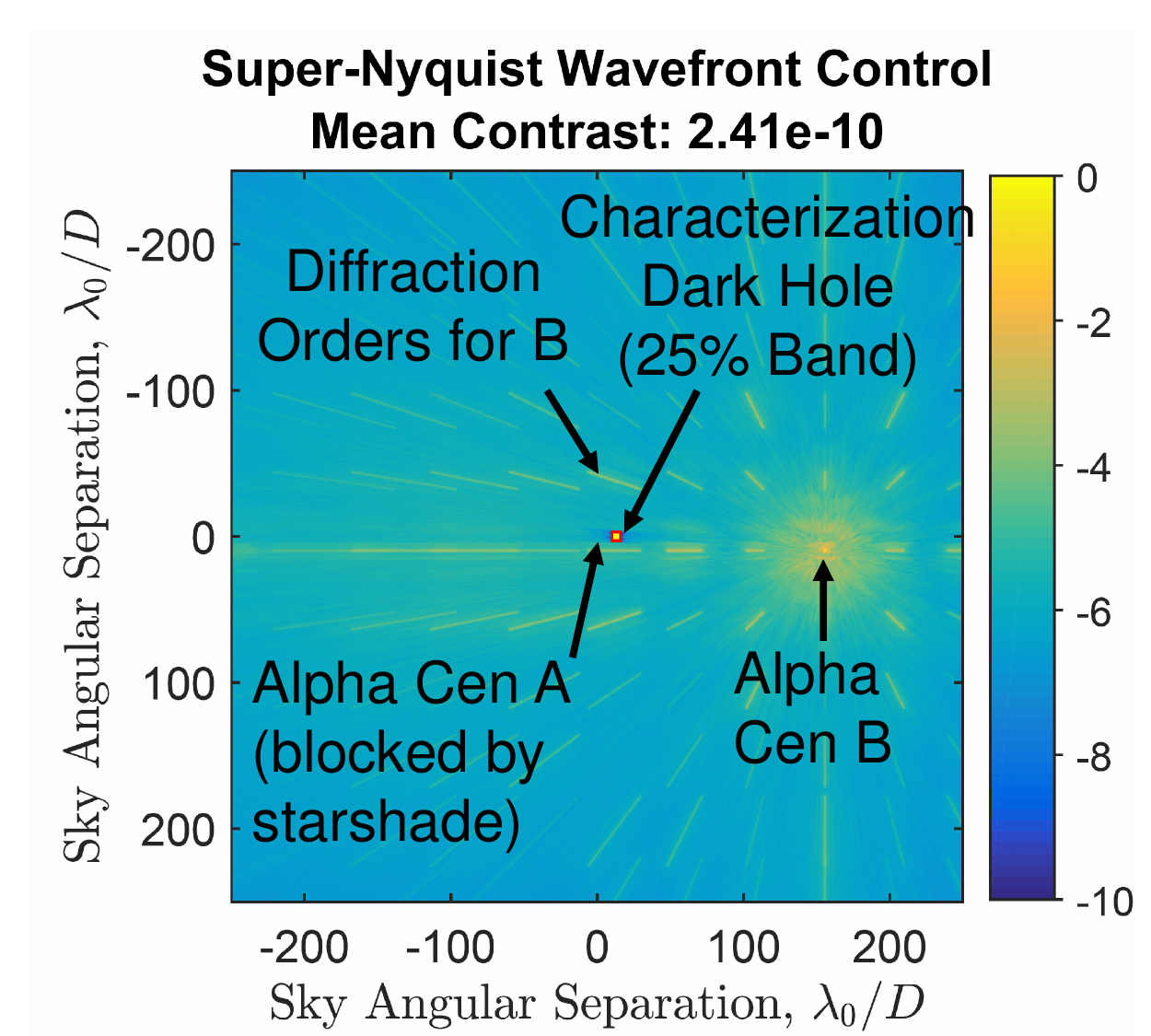}
}
\subfloat[]{
	\centering
	\label{fig:snwcObservation-characterization-zoomIn}
	\includegraphics[width  = 0.45\columnwidth, trim =0in 0in 0in 0in]{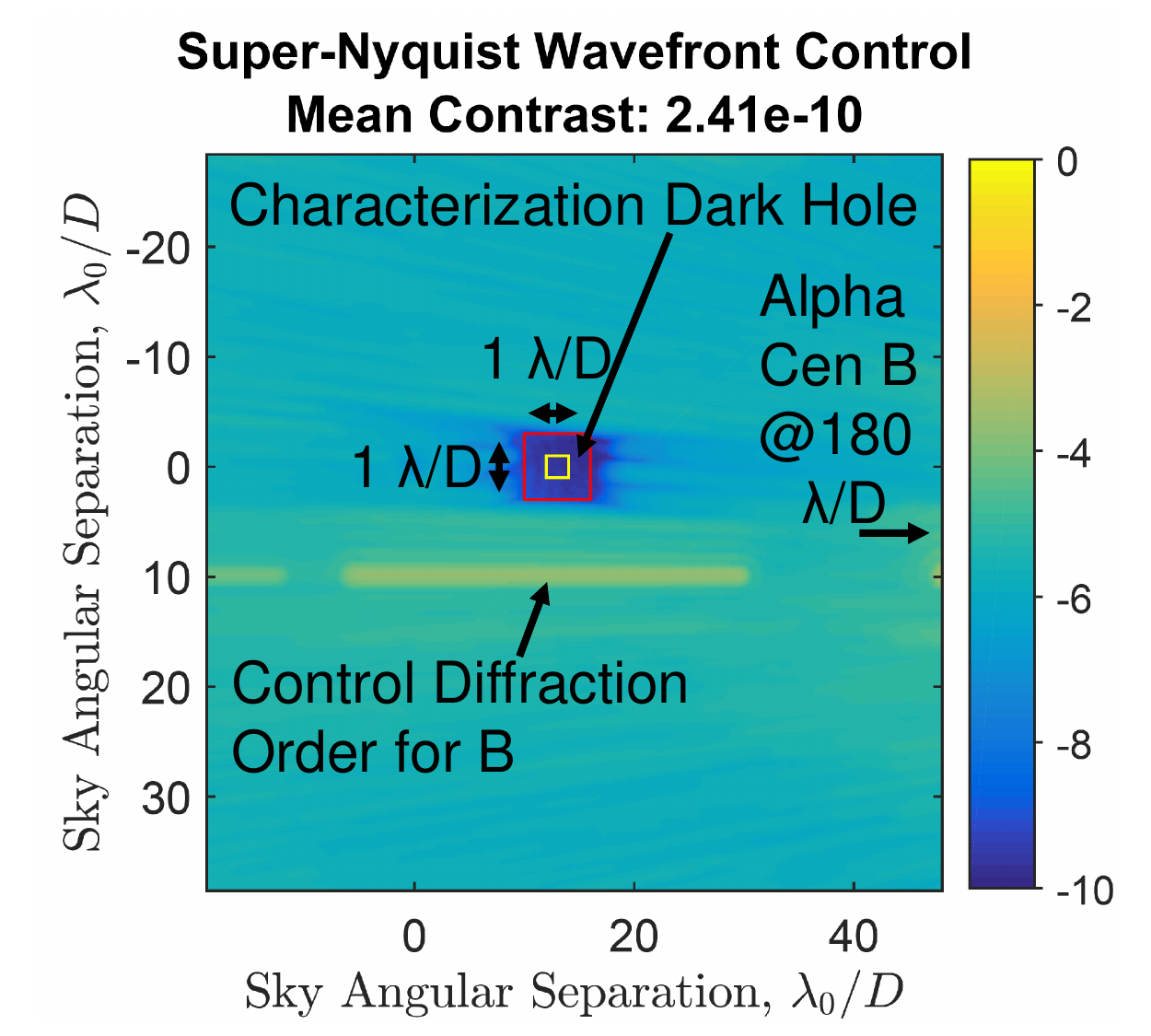}
}
\caption[Companion Leakage]
{\label{fig:snwcObservation} Observation strategy demonstration using on-axis starshade and SNWC to remove off-axis speckles follows two steps:  (1)  a larger $10\times10 \lambda/D$ dark hole with a 10\% bandwidth discovery region is first created, then (2) follow-up characterization observations are made with a smaller $1\times1 \lambda/D$ dark hole with a wider 25\% bandwidth. Figure panels show: \subref{fig:snwcObservation-discover-zoomOut} Zoom-out of discovery observation showing the binary star and diffraction orders in 10\% band. \subref{fig:snwcObservation-discover-zoomIn} Zoom-in of discovery observation showing the dark hole. \subref{fig:snwcObservation-characterization-zoomOut} Zoom-out of characterization observation showing the binary star and diffraction orders in 25\% band. \subref{fig:snwcObservation-characterization-zoomIn} Zoom-in of characterization observation showing the dark hole.}
\end{figure}

We consider a second simulation in Figure \ref{fig:snwcObservation} for the Alpha Centauri stars separated by 180 $\lambda/D$ to illustrate a possible observation strategy using this configuration. The observation strategy is as follows:
\begin{enumerate}
\item Generate a larger discovery region at smaller bandwidths. This would enable discovering new planets.
\item Follow-up with a smaller charecterization region at wider bandwidths. This would enable spectroscopic characterization of already-known exoplanets.
\end{enumerate}
In Figure \ref{fig:snwcObservation-discover-zoomOut} the zoomed-out field of view corresponding to the discovery observation is shown. A 10\% bandwidth is used to generate a dark hole with 10 $\lambda/D \times 10$ $\lambda/D$ size. The resulting zoomed-in field of view across the discovery dark hole near the elongated diffraction order used to control the speckles of Alpha Centauri B is clearly shown in Figure \ref{fig:snwcObservation-discover-zoomIn}. Achieved contrast across the discovery dark hole is $1.34 \times 10^{-10}$.

The follow-up observation configuration is shown in Figure \ref{fig:snwcObservation-characterization-zoomOut}. The diffraction orders are elongated further due to the increased 25\% bandwidth. SNWC using EFC optimized for broadband (by controlling 7 evenly spaced wavelengths across the band). To enable wider bandwidths the size of the dark hole can be traded provided that the exoplanet location is known a priori. In this case, the characterization dark hole is only $1 \lambda/D \times 1 \lambda/D$, although a small margin is added to minimize the effect of intermediate wavelengths not explicitly controlled. This resulting characterization dark hole is shown in Figure \ref{fig:snwcObservation-characterization-zoomIn}. Mean raw contrast achieved across the dark hole in 25\% broadband is $2.41 \times 10^{-10}$. 

An advantage of this method is that if a starshade is available using it on-axis provides deep suppression of the nearby star in wide bandwidths. As a result the impact of the WFIRST pupil is mitigated and hybrid operation with the CGI DM allows removal of off-axis leakage to enable deep raw contrasts at the $10^{-10}$ or deeper as demonstrated via simulated examples. The proposed observation strategy allows wider fields of view for discovery at the cost of bandwidth, and follow-on spectroscopic characterization can be achieved with smaller dark holes. The simulated example has shown a dark hole with 25\% bandwidth at large angular separations is possible (wider than the IFS bandwidth that is currently set at 20\%). Finally, this multi-star imaging scenario uses the starshade alignment as designed with the LOWFS sensor performing position sensing. To enable observation of binary stars with Super-Nyquist angular separations, the only requirement is that a diffraction grating be available at an appropriate location within the CGI (for example could take the form of a diffractive pupil \cite{Bendek13}). SNWC is otherwise a purely an algorithmic solution.

\section{Conclusions}  \label{sect:conc}

In this study, we have outlined and discussed the challenges associated with multi-star imaging due to the optical aberrations across the telescope optics giving rise to off-axis leakage. The options available to mitigate off-axis leakage were summarized and illustrated. Finally, multi-star imaging scenarios using a starshade with WFIRST were performed.

\begin{figure}[t!]
\centering
\includegraphics[width  = 1\columnwidth, trim = 0in 0.75in 0in 0in]{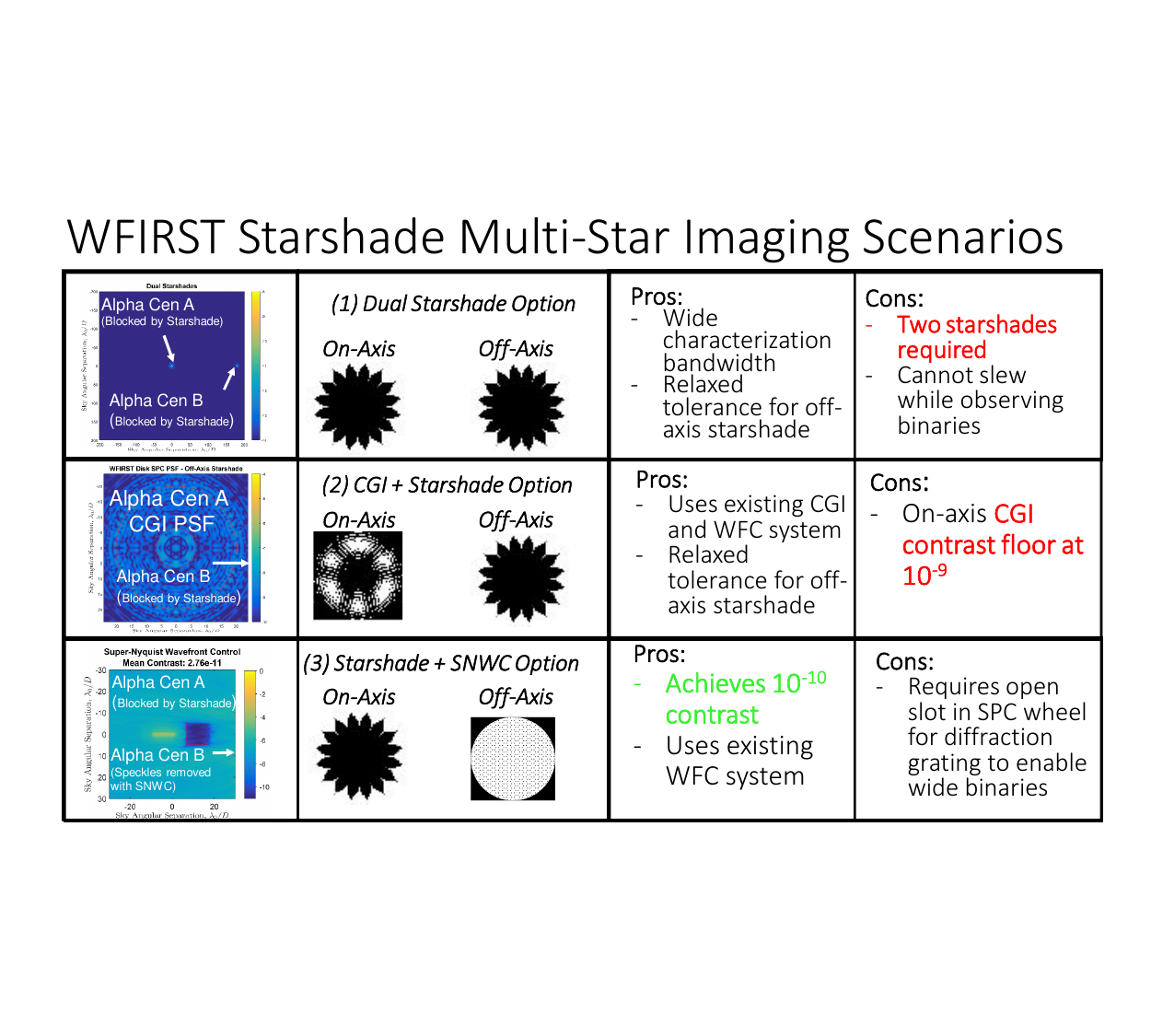}
\caption[Summary Table]
{\label{fig:starshadeMultiStarSummary} Summary of WFIRST scenarios for multi-star imaging using starshades in (1) dual mode or in hybrid combinations with wavefront control (WFC) with the starshade located (2) off-axis or (3) on-axis.}
\end{figure}

The multi-star imaging scenarios possible with a starshade operated with WFIRST are summarized in Figure \ref{fig:starshadeMultiStarSummary}. These include (1) a dual starshade option (with both on-axis and off-axis starshades), (2) the on-axis WFIRST CGI with an off-axis starshade, and (3) the on-axis starshade with SNWC removing off-axis leakage. Additionally, it is worth noting that a coronagraph with only a DM and a diffraction grating can also image multi-star system using MSWC \cite{Belikov17, Sirbu17}.

A rendezvous starshade probe with WFIRST enables the possibility of direct imaging of multi-star systems such as Alpha Centauri at deep raw contrast at the $10^{-10}$ level necessary for  direct imaging of rocky exoplanet. A particularly interesting option is hybrid operation of the on-axis starshade mitigating the raw contrast restrictions due to the WFIRST pupil combined with SNWC to enable off-axis leakage removal at wide angular separations using the already available DM onboard the WFIRST CGI. This study has shown through simulated demonstrations the feasibility of using a hybrid starshade and wavefront control system in the form of SNWC to enable multi-star imaging at wide spatial and spectral bandwidts. Future work will be focused on performing a sensitivity analysis to outline the capabilities for future hybrid systems such as WFIRST and HabEx to determine the trade-offs involved in spatial/spectral bandwidths and robustness to finite stellar sizes and dynamical disturbances.

\acknowledgments

This work was supported in part by the National Aeronautics and Space Administration's Ames Research Center, as well as the NASA Astrophysics Research and Analysis (APRA) program through solicitation NNH13ZDA001N-APRA at NASA's Science Mission Directorate. Any opinions, findings, and conclusions or recommendations expressed in this article are those of the authors and do not necessarily reflect the views of the National Aeronautics and Space Administration. DS was supported for part of this work by a NASA Postdoctoral Program fellowship. This research has made use of the Washington Double Star Catalog maintained at the U.S. Naval Observatory.


\bibliography{refs}   
\bibliographystyle{spiebib}   

\end{document}